\documentclass[secnumarabic, reprint, superscriptaddress, tightenlines, amsmath,amssymb, aps, prl, nobibnotes,
]{revtex4-1}

\usepackage{xfrac}
\usepackage{float}
\usepackage{natbib}
\usepackage{graphicx,caption,subcaption,xcolor}
\usepackage[colorlinks=true,citecolor=blue,
linkcolor=blue,urlcolor=blue]{hyperref}
\usepackage[utf8]{inputenc} 
\usepackage[english]{babel}
\usepackage[OT2, T1]{fontenc}
\usepackage{upquote}
\usepackage{blindtext}
\usepackage{mathtools}
\usepackage [autostyle, english = american]{csquotes}
\usepackage{bigints}

\usepackage{multirow}

\usepackage{amssymb}% http://ctan.org/pkg/amssymb
\usepackage{pifont}% http://ctan.org/pkg/pifont
\newcommand{\cmark}{\ding{51}}%
\newcommand{\xmark}{\ding{55}}%
\allowdisplaybreaks

\usepackage{tikz}
\usepackage[compat=1.1.0]{tikz-feynman}
\usetikzlibrary{positioning,calc}

\usepackage{amssymb}% http://ctan.org/pkg/amssymb
\usepackage{pifont}% http://ctan.org/pkg/pifont

\usepackage{makecell}

\begin{document}

\title{Subgap Bound States 
%and Spectroscopic Signatures\\  
from Dynamical Impurities}

\author{Joshuah T. Heath}
\email{joshuah.heath@mu.ie}

\affiliation{Department of Physics, Science Building, Maynooth University, Maynooth, Co. Kildare W23 F2H6, Ireland}
\affiliation{Hamilton Institute, Eolas Building, Maynooth University, Maynooth, Co. Kildare W23 A3HY, Ireland}
\affiliation{Nordita, Stockholm University and KTH Royal Institute of Technology, Hannes Alfvéns väg 12, SE-106 91 Stockholm, Sweden}
\affiliation{Department of Physics, University of Connecticut, Storrs, Connecticut 06269, USA}

\date{\today}

\begin{abstract}
\noindent We find that a single non-magnetic dynamical impurity 
(e.g., a low-frequency localized "vibron" mode) 
induces  a subgap bound state in an s-wave superconductor. A closed-form solution for the bound state energy is found as a function of the vibron frequency in the elastic scattering limit, with the salient features of the bound state energies unaffected by inelastic scattering processes in the sub-THz regime. In addition to subgap features in the low-frequency local density of states, such impurities result in real space regions of reduced spectroscopic weight outside the gap peak.
~%Such impurities result in real space regions characterized by spectroscopic features outside the gap peak and .
%These findings indicate the unexpected role of dynamic defects, like two level systems, ubiquitously seen in superconductors, as the source of subgap quasiparticle bound states.  
%These findings suggest that dynamical defects in superconductors can be a source of subgap quasiparticle bound states.
%real-space characterized by a low-energy density of states.
%at high impurity concentration.
%, and suggest that dynamical impurities may serve as a      
\end{abstract}

\pacs{1}

\maketitle

\indent {\it Introduction --} The influence of disorder on  superconductivity is a broad topic that spans over half a century~\cite{AG1959, RickayzenBook,DeGennes1966,Abrikosov1969Feb,Finkelstein1994Mar,Balatsky2006May}. Microscopic defects can suppress the gap in cases beyond s-wave pairing, and are thus often utilized as a probe of unconventional superconductivity. This is well known in SrRuO$_4$~\cite{Mackenzie1998Jan,Mackenzie2003May}, where both magnetic and non-magnetic impurities can suppress the local superconducting gap~\cite{Millis1988Apr}. 
%Indeed, often impurity-induced suppression of superconductivity is treated as a strong signature of non-BCS type pairing. 
In contrast, the suppression of superconductivity via impurities in s-wave superconductors is highly dependent on the individual impurity's potential.
%their 
%Broadly, defects in superconductors are often classified according to their 
%scalar potential vs. magnetic character. 
The scalar impurity potential is typically modelled as a local change in the chemical potential, whereas a magnetic impurity %enters as a $S \cdot \sigma_z \otimes \tau_3$ term. It has been long established that, in an s-wave superconductor, a single magnetic impurity 
has a more complex structure that induces a quasiparticle bound state within the gap (a subgap state via local exchange interaction) known as a Yu-Shiba-Rusinov (YSR) state~\cite{Luh1965,Shiba1968Sep,Rusinov}. Note that the static impurity does not lead to the formation of a quasiparticle bound state in conventional superconductors, and thus s-wave BCS superconductivity remains robust in the presence of static, non-magnetic disorder~\cite{Anderson1959Sep}. 

In this Letter, we extend the classification of defects in superconductors and address the question of subgap  states induced by local dynamical defects, which we model as local variations of the ground state energy. One example of such a dynamical impurity is a low-frequency mode localized in real space, dubbed a "vibron"~\cite{Zhu2004Jul,Fransson2010May}. Vibrons are ubiquitous in disordered materials, and 
have recently been proposed as a candidate for microscopic two-level system (TLS) defects commonly found in amorphous solids~\cite{Phillips1987Dec,W.Anderson1972Jan,ZacconeBook}, e.g. in superconductors~\cite{Kristen2024May,He2025Jun,Grankin2026Jul} and the native oxides of substrate-metal interfaces~\cite{DuBois2013Feb,Muller2019Oct,Tyner2025Oct}. Most recently, vibrons have been proposed to induce localized "hotspots" of Josephson energy at the tunnel junction barrier~\cite{Heath2026Jan}, and may be used to engineer higher $T_c$ in the thin-film limit~\cite{Samoilenka2020Apr,Heath2026May}. 

 \begin{figure}[t]
\includegraphics[width=1\columnwidth]{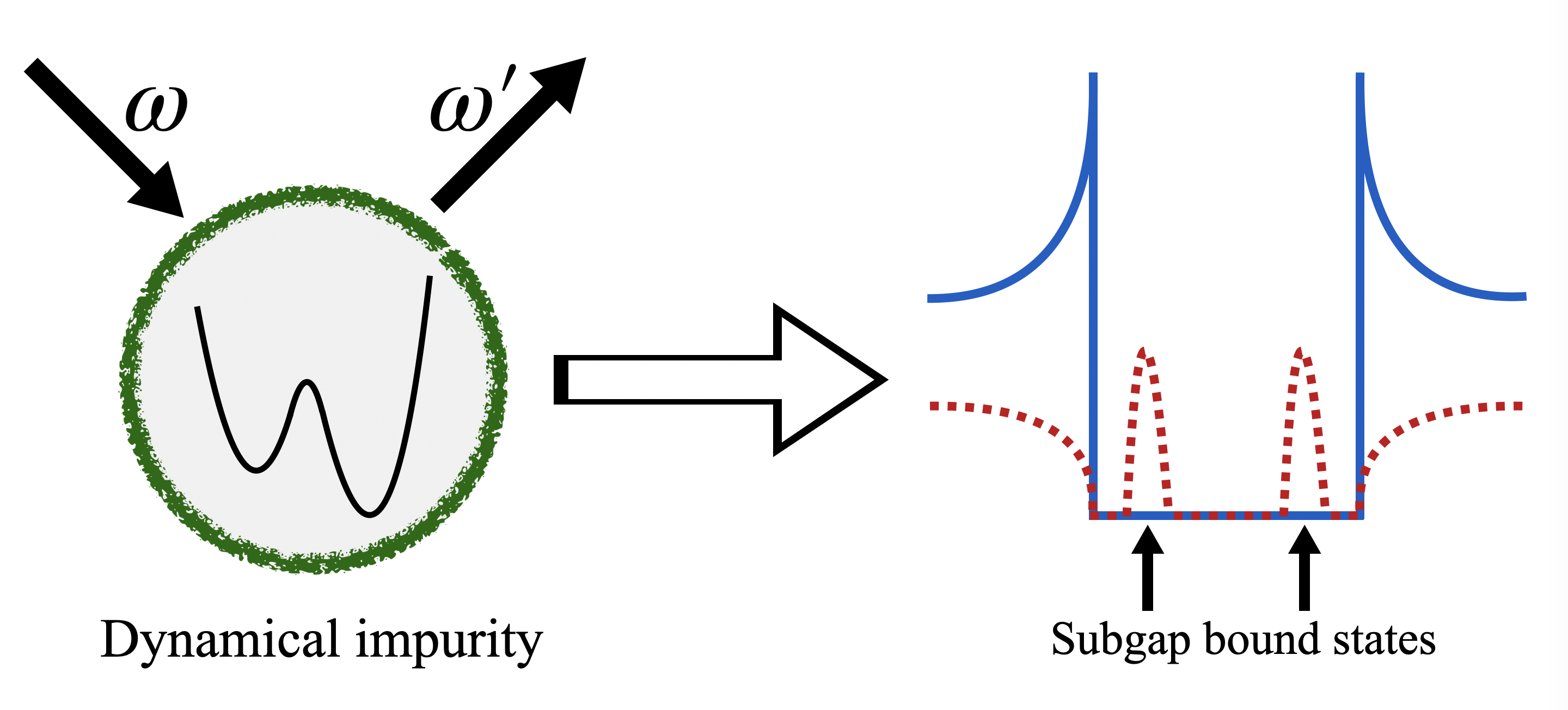}
\caption{\small (Left)  
An electron scattering off a low-frequency dynamical impurity, which in this context is represented by a two-level system (TLS) double-well potential. (Right) A schematic of the local superconducting density of states (LDOS) in the absence of a dynamical impurity (blue, solid) and in the presence of such an impurity (red, dotted). Subgap bound states emerge for $|\omega|<\Delta_0$, while the LDOS is suppressed for $|\omega|>\Delta_0$. 
%The bound state energy is expressed in terms of the local electron-boson coupling $\lambda_V(\omega_V,\epsilon)$, which is a function of the vibron frequency $\omega_V$ and the electron-boson spectral function width $\epsilon$.
\label{fig:firstpage}}
\end{figure}

We find that non-magnetic {\it dynamical} defects in s-wave superconductors, regardless of their microscopic origin, induce i) bound states for frequencies below the gap and ii) robust spectroscopic features for frequencies above the gap. The frequency $\omega^*$ of these subgap bound states can be exactly solved for in the limit of elastic scattering. Taking the example of a local boson/vibron, we find that  
%\newpage
\vspace{-00mm}
\begin{align}
    \omega^*/\Delta_{\textrm{hom}}=\pm \bigg\{
\dfrac{1 -\pi^2\lambda_V^2(\omega_V,\epsilon)}{1+\pi^2\lambda_V^2(\omega_V,\epsilon)}
    \bigg\},\label{Eqn1}
\end{align}

\noindent where $\Delta_{\textrm{hom}}$ is the $T=0$ BCS gap in the homogeneous (i.e., defect-free) material and $\lambda_V(\omega_V,\,\epsilon)$ is the local electron-boson coupling induced by a dynamical impurity with frequency $\omega_V$ and electron-boson spectral function of width $\epsilon$. Our work thus suggests that a severely inhomogeneous energy landscape destabilizes superconductivity regardless of the gap anisotropy (see Table I).

% As a whole, our work suggests that 

% These dynamical defects represent a new class of local scatterers, as their BdG structure is different from local static potential scattering $\tau_3$ and local magnetic scattering $S \cdot \sigma_z \otimes \tau_3$ (see Table I)
%In addition, we consider the more realistic treatment of off-shell contributions to impurity scattering, and show that they do not have a significant effect on the salient features of Eqn.~\eqref{Eqn1}.
%, suggesting that our main hypothesis remains valid when energy is exchanged between the vibron and incident electron. 
%From the $T$-matrix, we find out-of-gap signatures 
%in the local superconducting density of states (LDOS) 
%may be detectable via STM measurements of the LDOS.
%detectable by inelastic electron tunnelling spectroscopy (IETS). 
%Such dynamical defects induce robust oscillation features for frequencies above the gap, resulting in experimentally-relevant real-space modulation of the local density of states (LDOS) not seen in static impurities.
%; and 
% these dynamical defects induce  persistent oscillation features in the local Density of states (LDOS); v) electron scattering off local dynamical impurity may experience a shift of energy~\cite{Balatsky2006May} and result in inelastuc electron tunneling spectroscopy (IETS) features most prominently seen outside the coiherence peaks. 
\begin{table}[t]
\centering
\hspace{-5mm}\begin{tabular}{|c|c|c|c|}
\hline\
Impurity class & pole s-wave & pole p/d-wave & $\mu_m\not=0$ \\
\hline\hline
Non-magnetic $\tau_3$ & {\color{red}\xmark} & \cmark & {\color{red}\xmark} \\
\hline
Magnetic $\tau_0 \otimes {\bf S}\cdot {\bf \sigma}$ & \cmark & \cmark  & \cmark\\
\hline
Dynamical $\tau_0$ & \cmark & \cmark & {\color{red}\xmark} \\
\hline
\end{tabular}
\caption{Classification of different impurities by their Nambu matrix structure in the BdG Hamiltonian. We model dynamical impurities as some energy inhomogeneity, which manifests as an effective $\tau_0$ potential. Such an impurity potential is unique in that it induces a pole in the s-wave superconductor in the absence of a magnetic moment.}
\label{tab:example}
\end{table}
{\it Bound states from electron-vibron scattering--} In a homogeneous (i.e., impurity-free) many-body system, the local interacting Green's function $\mathcal{G}_{\textrm{hom}}(i\omega_n;\,{\bf r})$ is related to the local free-electron Green's function $\mathcal{G}_{0}(i\omega_n;\,{\bf r})$ via the local self-energy $\Sigma(i\omega_n)$. For our purposes, we are concerned with impurities in superconductors, and thus we consider a Nambu-Gor'kov structure for the propagator given by $\mathcal{G}_0(i\omega_n;\,{\bf r})=G_0(i\omega_n;\,{\bf r})\tau_0+F_0(i\omega_n;\,{\bf r})\tau_1$, where $G_0(i\omega_n;\,{\bf r})$ is the normal Green's function, $F_0(i\omega_n;\,{\bf r})$ is the anomalous Green's function, and $\tau_i$ are Pauli matrices in particle-hole space. Throughout this paper, propagators written in cursive are assumed to be in the Nambu matrix structure.

In the presence of a single impurity, we utilize the $T$-matrix approximation~\cite{Hirschfeld1986Jul,Hirschfeld1993Aug,Hotta1993Dec,Ziegler1996Apr,Hussey2002Dec,Bruus2004Sep,Balatsky2006May,Bena2016Mar}. Within this approximation, the full local Green's function $\mathcal{G}(i\omega_n,\,i\omega_m;\,{\bf r},\,{\bf r}')$ is given by the following equation:
%, and shown in Fig.~\ref{fig:Feynman}(b):

% In the presence of a single impurity, we must take into account
% translational symmetry breaking induced by the electron scattering off that impurity, and therefore we utilize the $T$-matrix approximation~\cite{Hirschfeld1986Jul,Hirschfeld1993Aug,Hotta1993Dec,Hussey2002Dec,Balatsky2006May}. Within this approximation, the full local Green's function $\mathcal{G}(i\omega_n,\,i\omega_m;\,{\bf r},\,{\bf r}')$ is given by the following equation, and shown in Fig.~\ref{fig:Feynman}(b):

\begin{align}
&\hspace{17.9mm}\mathcal{G}(i\omega_n,\,i\omega_m;\,{\bf r},\,{\bf r}')=\mathcal{G}_{\textrm{hom}}(i\omega_n;\,{\bf r}-{\bf r}')\delta_{n,m}\notag\\
&+\mathcal{G}_{\textrm{hom}}(i\omega_n;\,{\bf r}-{\bf r}_i)\mathcal{T}(i\omega_n,\,i\omega_m;\,{\bf r}_i)\mathcal{G}_{\textrm{hom}}(i\omega_m;\,{\bf r}_i-{\bf r}'),
\end{align}
\noindent where ${\bf r}_i$ is the real-space position of the single impurity and the $T$-matrix $\mathcal{T}(i\omega_n,\,i\omega_m;\,{\bf r}_i)$ is defined by the following self-consistent Lippmann–Schwinger equation~\cite{Bruus2004Sep}:
\begin{align}
    &\hspace{23mm}\mathcal{T}(i\omega_n,\,i\omega_m;\,{\bf r}_i)=\mathcal{V}(i\omega_n,\,i\omega_m;\,{\bf r}_i)\notag\\
    &+\sum_{\ell}\mathcal{V}(i\omega_n,\,i\omega_\ell;\,{\bf r}_i)\mathcal{G}_{\textrm{hom}}(i\omega_\ell;\,{\bf r}_i)\mathcal{T}(i\omega_\ell,\,i\omega_m;\,{\bf r}_i),\label{TMatrix}
\end{align}
\noindent where $\mathcal{V}(i\omega_n,\,i\omega_m;\,{\bf r}_i)$ is a general form for the impurity potential induced by an impurity at position ${\bf r}_i$. 

In Eqn.~\ref{TMatrix}, we assume the potential and the $T$-matrix are independent of the wave vector ${\bf k}$, with the sum over intermediate momenta absorbed into the definition of the local homogeneous Green's function. Also note that, for a dynamical impurity, the electron may transmit or absorb energy when in contact with the defect, thereby leading to inelastic scattering where the ingoing and outgoing energies are not equal. In the most general form of the local $T$-matrix given in Eqn.~\ref{TMatrix}, we therefore include a sum over a finite "ladder" of intermediate energies induced by repeated electron-vibron scattering. If we assume the case of elastic scattering (i.e., $n=m$), the sum over intermediate energies collapses and an analytical equation for the $T$-matrix may be found.
%For the static impurity 

 \begin{figure}[t]
\hspace{-5.75mm}\includegraphics[width=1.05\columnwidth]{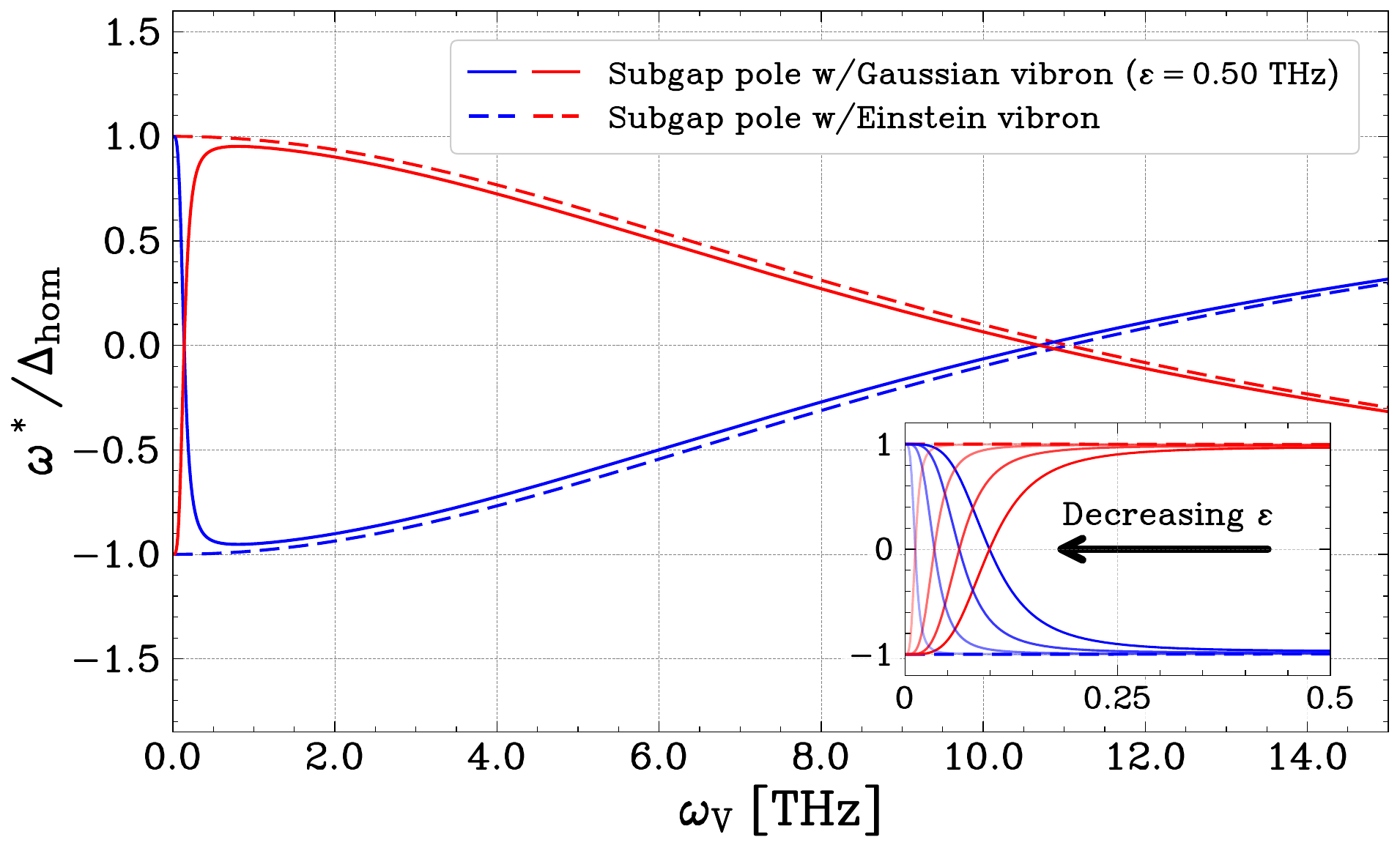}
\caption{\small Dependence of the subgap pole structure $\omega^*$ on the vibron frequency $\omega_V$. The bulk electron-phonon coupling strength is taken to be $\lambda_0=0.43$ (i.e., for Al).
%, but the features of the pole structure are agnostic to the specific s-wave superconductor. 
If modelled as an Einstein boson (dashed red and black lines), the vibron frequency sits near the BCS gap $\Delta_0$ in the static limit before decreasing to zero frequency around $\omega_V\sim 11$ THz and then slowly returning to $\Delta_0$. The Gaussian vibron (characterized by a spectral function $\alpha^2 F(\nu)$ with a non-zero width $\epsilon$) has very similar behavior, although experiences an additional low-frequency crossing in the GHz regime. (Inset) The GHz structure of the pole for decreasing $\epsilon$. As $\epsilon\rightarrow 0$, the first crossing of the poles approaches $\omega_V=0$, recovering the Einstein boson result.
\label{fig:boundstatepoles}}
\end{figure}

%  \begin{figure}[t]
% \hspace{-10.75mm}\includegraphics[width=1.05\columnwidth]{subgappole.pdf}
% \caption{\small test.
% \label{fig:boundstatepoles}}
% \end{figure}

For a static non-magnetic impurity, the scattering potential will have the same Nambu form as the chemical potential, and thus we may call such a defect a $\tau_3$-impurity~\cite{Balatsky2006May}. Similarly, a local inhomogeneity of the pair potential may result in a $\tau_1$ impurity~\cite{Nunner2005Oct,Zhu2006Oct,Bespalov2016Mar,Morin2026Jul}. For a purely dynamical impurity (i.e., ignoring the effects of the density modulation), we should instead have a particle-hole symmetric frequency/energy shift, and thus the vibron may be considered to be a $\tau_0$ impurity (see Table I). For simplicity, we assume that the static contribution (i.e., $\tau_3$ potential) is negligible compared to the $\tau_0$ potential, and thus we have a purely dynamical impurity.

 \begin{figure}[t]
\hspace{-5.75mm}\includegraphics[width=1.05\columnwidth]{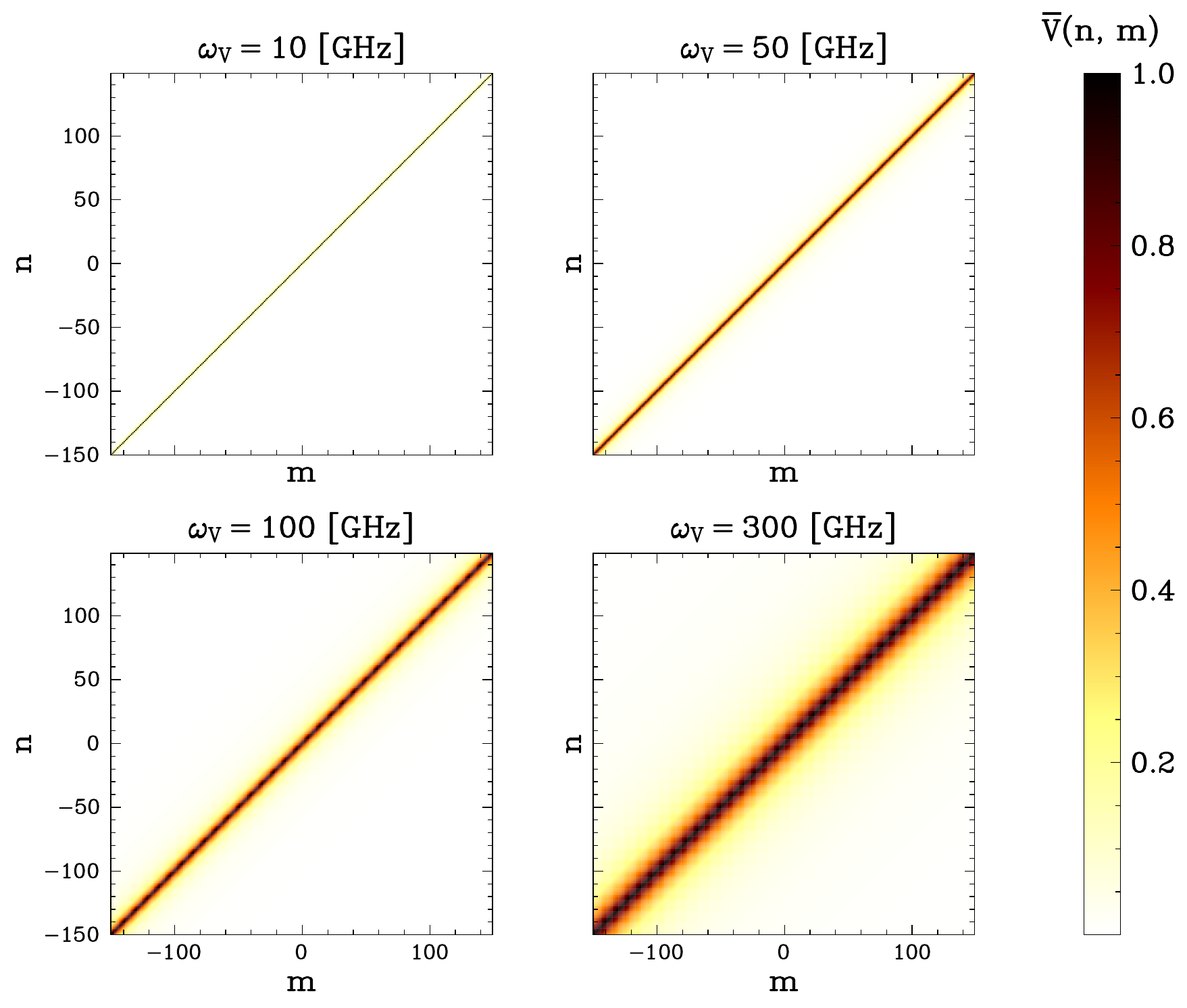}%{V_inelastic.pdf}
\caption{\small The normalized scattering potential ${V}(n,\,m)\equiv \mathcal{V}(i\omega_n,\,i\omega_m;\,{\bf r}_i)$ for the dynamical impurity normalized to one, plotted a function of the Matsubara indices $n$ and $m$ for several different values of the vibron frequency $\omega_V$. Perfect elastic scattering implies that all weight is on the $n=m$ scattering, and thus the elastic scattering limit manifests as a diagonal on the above plots. Upon increasing the vibron frequency, more elements off the diagonal of the potential become non-zero, resulting in an increasing dominance of inelastic scattering events. The elastic approximation remains reasonable below $\omega_V=100$ GHz.
\label{fig:V}}
\end{figure}

Our main conclusions (namely, i and ii in the introduction) are agnostic to the form of the $\tau_0$ potential. While a $\tau_0$ may be argued solely by imposing parity and time-reversal symmetry~\cite{Slager2015Aug,He2025Jun}, to make material-specific estimates we will assume a form of the potential assuming an electron interacting with a single dynamical defect; namely, $\mathcal{V}(i\omega_n,\,i\omega_m;\,{\bf r}_i)\equiv V_0(i\omega_n,\,i\omega_m;\,{\bf r}_i)\tau_0$, where $V_0(i\omega_n,\,i\omega_m;\,{\bf r}_i)= g^2 D(i\omega_n-i\omega_m)\delta_{{\bf r}_i,0}$, $g$ is the bare electron-vibron coupling, and $D(i\omega_n-i\omega_m)$ is the vibron propagator. This is the undressed potential induced by the simultaneous exchange/absorption of a localized boson at ${\bf r}_i=0$, and is taken as the lowest-order approximation in the $T$-matrix. A similar approach is taken in the study of electron-electron scattering via self-interaction induced by a bosonic condensate~\cite{Nozieres1985May,Lipavsky2008Dec,Sopik2011Sep}, where the effective potential includes the interaction vertices and thus should scale as a local energy shift. In our case, the local exchange of a bosonic mode is mediated by a dynamical defect, and thus we are only concerned with single-fermion scattering in the $T$-matrix. In this way, our estimate of $V_0$ is similar to a polaron-like interaction~\cite{Marsiglio1991Mar,Combescot1995May,Hohenadler2007,Dai2025Dec,Chubukov2026Apr,Supp}, and is therefore analogous to scattering off of an isotopic impurity~\cite{Denisov2024Jul} or an Anderson-Holstein impurity~\cite{Hewson2001Dec,Al-Eryani2026Apr}.

Taking a general vibronic mode of frequency $\omega_V$ described by a Lorentzian electron-boson spectral function $\alpha^2 F(\nu)$ with width $\epsilon$~\cite{Marsiglio2020Jun}, the potential is given by
\begin{align}
    &V(i\omega_n,\,i\omega_m;\,{\bf r}_i)=\dfrac{2g^2}{\pi}\int_0^{2\omega_V}d\nu \bigg\{\dfrac{\nu}{(\omega_n-\omega_m)^2+\nu^2}\bigg\}\notag\\
    &\phantom{-\tau_0\dfrac{2g^2}{\pi}\int_0^{2\omega_V}d\nu\cdot\cdot\quad\,\,\,\,\,}\times \bigg\{\dfrac{\epsilon}{(\nu-\omega_V)^2+\epsilon^2}-\dfrac{\epsilon}{\omega_V^2+\epsilon^2}\bigg\}.
    %+\tau_3V(0;\,{\bf r}_i)
\end{align}

\begin{table}[t]
\centering
\begin{tabular}{|c|c|c|c|c|c|c|}
\hline
Material &
$\lambda_0$ &
\makecell[c]{$\omega_{\ln}$ \\ {[THz]}} &
\makecell[c]{$\Delta_{\textrm{hom}}$ \\ {[GHz]}} &
\makecell[c]{\rule{0pt}{2.6ex}$\omega^*$ [GHz] \\ $\omega_V=15$ GHz \\ $\epsilon$ minimized} &
$\dfrac{\omega^*}{\Delta_{\textrm{hom}}}$ &
Source \\
\hline\hline
NbN & 1.1 & 5.6 & 689 & 688.3 & 0.999 & \cite{Beck2011Oct,Noat2013Jul,Babu2019Mar}\\ \hline
%TiN & - & - & - & - & - & -\\ \hline
Nb & 1.1 & 3.2 & 363 & 361 & 0.993 & \cite{Butler1977Oct,Arnold1980Aug,Carbotte1990Oct}\\ \hline
%Re & - & - & - & - & - & -\\ \hline
Al & \,0.43\, & 6.2 & 44 & 43 & 0.988 & \cite{Leung1976,Carbotte1990Oct}\\ \hline 
Ta & \,0.69\, & 2.7 & 174 & 154 & 0.886 & \cite{Carbotte1990Oct}\\ \hline
%$\alpha$-Ga & - & - & - & - & - & -\\ \hline
Pb & 1.55 & 1.17 & 338.5 & 273.4 & 0.808 & \cite{Carbotte1990Oct}\\ \hline
Sn & 0.72 & 2.0 &  \,\,146.5\,\, & 117.6 & 0.802 & \cite{Carbotte1990Oct}\\ \hline
In & 0.81 & 1.4 &  131 & 79 & 0.601 & \cite{Carbotte1990Oct}\\ \hline
Hg & \,\,1.62\,\, & 0.6 &  201 & 41.3 & 0.206 & \cite{Carbotte1990Oct}\\ \hline
% Material & $\lambda_0$ & $\omega_{\ln}$ [THz] & $\Delta_0$\,[GHz] &
% \makecell[c]{\\[.1pt] Min. $\omega^*$ [GHz] \\ $\omega_V=25$ GHz \\ $\epsilon$ minimized} & $\omega^*/\Delta_0$ \\
% \hline\hline
% NbN & d & r & 590 & 584 & 0.99\\ \hline
% TiN & d & r & 165 & 164.7 & 0.99\\ \hline
% Nb & \,\,1.1\,\, & r & 340 & 326 & 0.96\\ \hline
% Re & d & r & 62 & 59 & 0.95\\ \hline
% Al & \,0.43\, & 6.2 & 44 & 37 & 0.85\\ \hline %DONE; source Carbotte review
% Ta & \,0.69\, & 2.7 & 174 & 118 & 0.71\\ \hline
% $\alpha$-Ga & d & r & 40 & 26 & 0.65\\ \hline
% Hg & d & r &  190 & 58 & 0.30 \\ \hline
% Sn & d & r &  165 & 26 & 0.17 \\ \hline
% In & d & r &  125 & 6.5 & 0.05\\ \hline
\end{tabular}
\caption{\small 
%*Please  add the column that lists g - electron vibron coupling constant assumed in these estimates*.
Estimates for the minimally-expected bound-state frequency $\omega^*$ for several materials, assuming a dynamical impurity at $\omega_V=15$ GHz (i.e., the "worst-case" scenario for a vibron/TLS with fixed frequency in the low GHz range). For each material, the homogeneous electron-phonon coupling $\lambda_0$, homogeneous BCS gap $\Delta_{\textrm{hom}}$, and log average frequencies $\omega_{\ln}$ used in our estimation are given and sourced. To minimize $\omega^*$, we want to maximize the vibron frequency $\omega_V$ and minimize the vibron spectral width $\epsilon$.
%$\omega_V$ and minimize $\epsilon$.
%We choose physically realistic values of the vibron frequency $\omega_V$ and spectral width $\epsilon$ such that subgap bound states are most pronounced; i.e., closest to zero energy. 
%We take $\omega_V = 15$ GHz, which is around the typical upper limit for TLS frequencies in superconducting quantum devices. 
The spectral width $\epsilon$ is minimized such that the enhanced electron–boson coupling remains just within the regime of validity of Eliashberg theory~\cite{Yuzbashyan2022Aug} over all $\omega_V$. 
%*WE NEED ALSO COLUMN FOR ASSUMED ELECTRON PHONON VALUES. \cite{PhysRevB.92.121404}
% JTH: Will add this
For materials with strong covalent bonding like niobium nitride, vibrons in the GHz regime are expected not to form detectable subgap bound states. However, for materials with weaker metallic bonding such as indium and mercury, non-negligible low-frequency bound states form at much lower vibron frequencies.
}
\label{tab:Materials}
\end{table}

%  \begin{figure}[t]
% \hspace{-5.75mm}\includegraphics[width=1.05\columnwidth]{V_inelastic.pdf}%{V_inelastic.pdf}
% \caption{\small The normalized scattering potential ${V}(n,\,m)\equiv \mathcal{V}(i\omega_n,\,i\omega_m;\,{\bf r}_i)$ for the dynamical impurity normalized to one, plotted a function of the Matsubara indices $n$ and $m$ for several different values of the vibron frequency $\omega_V$. Perfect elastic scattering implies that all weight is on the $n=m$ scattering, and thus the elastic scattering limit manifests as a diagonal on the above plots. Upon increasing the vibron frequency, more elements off the diagonal of the potential become non-zero, resulting in an increasing dominance of inelastic scattering events. The elastic approximation remains reasonable below $\omega_V=100$ GHz.
% \label{fig:V}}
% \end{figure}

\noindent In the limit of $\epsilon\rightarrow 0$, we obtain the Einstein phonon result, while non-zero $\epsilon$ ensures $\omega_V\rightarrow 0$ results in the electron-boson coupling constant $\lambda(\omega_V\rightarrow 0;\,\epsilon)\rightarrow 0$ (as expected in the static limit)~\cite{Supp}.

%%%%%%push to appendix
%\noindent A plot of the dimensionless electron-vibron coupling $\lambda(\omega_V)=N(0)\mathcal{V}(i\omega_n-i\omega_m;\,{\bf r}_i)$ at the impurity site ${\bf r}_i$ is shown in Fig.~\ref{fig:lambda} as a function of the vibron frequency $\omega_V$, assuming a native phonon frequency (i.e., Einstein frequency) of aluminium. Note that, as $\epsilon\rightarrow 0$, we obtain the result of an Einstein phonon, i.e. $\alpha^2 F_V(\nu)=N(0)g^2\delta(\nu-\omega_V)$. For such a system, we see a directly inverse relationship between the vibron frequency $\omega_V$ and the the coupling $\lambda(\omega_V)$ (blue dashed line in Fig.~\ref{fig:lambda}). Note that such a result becomes unphysical as $\omega_V\rightarrow 0$: in the DC limit, the electron-vibron coupling should go down to zero. We therefore include a small width $\epsilon$ to the vibronic $\alpha^2 F_V(\nu)$, which reproduces the correct result of $\lambda_V(\omega_V,\,\epsilon\not=0)\rightarrow 0$ in the limit of vanishing vibron frequency.

 \begin{figure}[t]
\hspace{-10mm}\includegraphics[width=1.0\columnwidth]{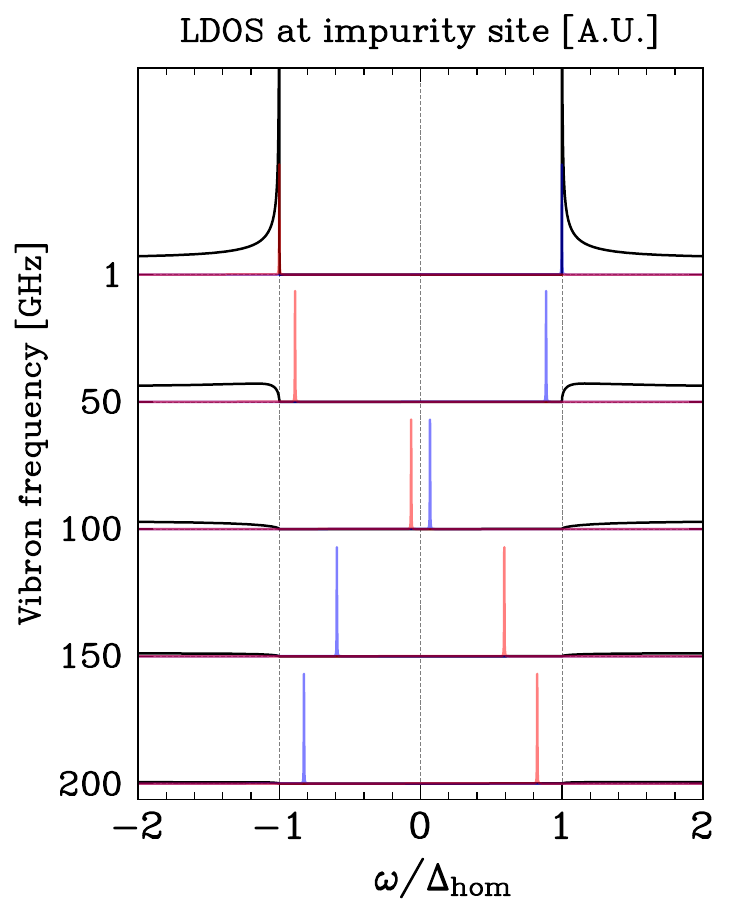}
\caption{\small The local density of states (LDOS) at the impurity site ${\bf R}_i={\bf r}-{\bf r}_i=0$ for several different vibron frequencies. The poles of the $T$-matrix are represented as blue and red delta functions centered at the bound state frequencies $\omega^*$. For $\omega_V$ below $10$ GHz, the subgap poles sit close to the BCS gap $\pm\Delta_0$. As the vibron frequency is increased, the subgap poles move inward to $|\omega/\Delta_{\textrm{hom}}|<1$ and cross, while the LDOS for $|\omega/\Delta_{\textrm{hom}}|>1$ is depleted as spectral weight moves towards the poles.
\label{fig:waterfall}}
\end{figure}

 \begin{figure}[t]
\hspace{-7.95mm}\includegraphics[width=1.1\columnwidth]{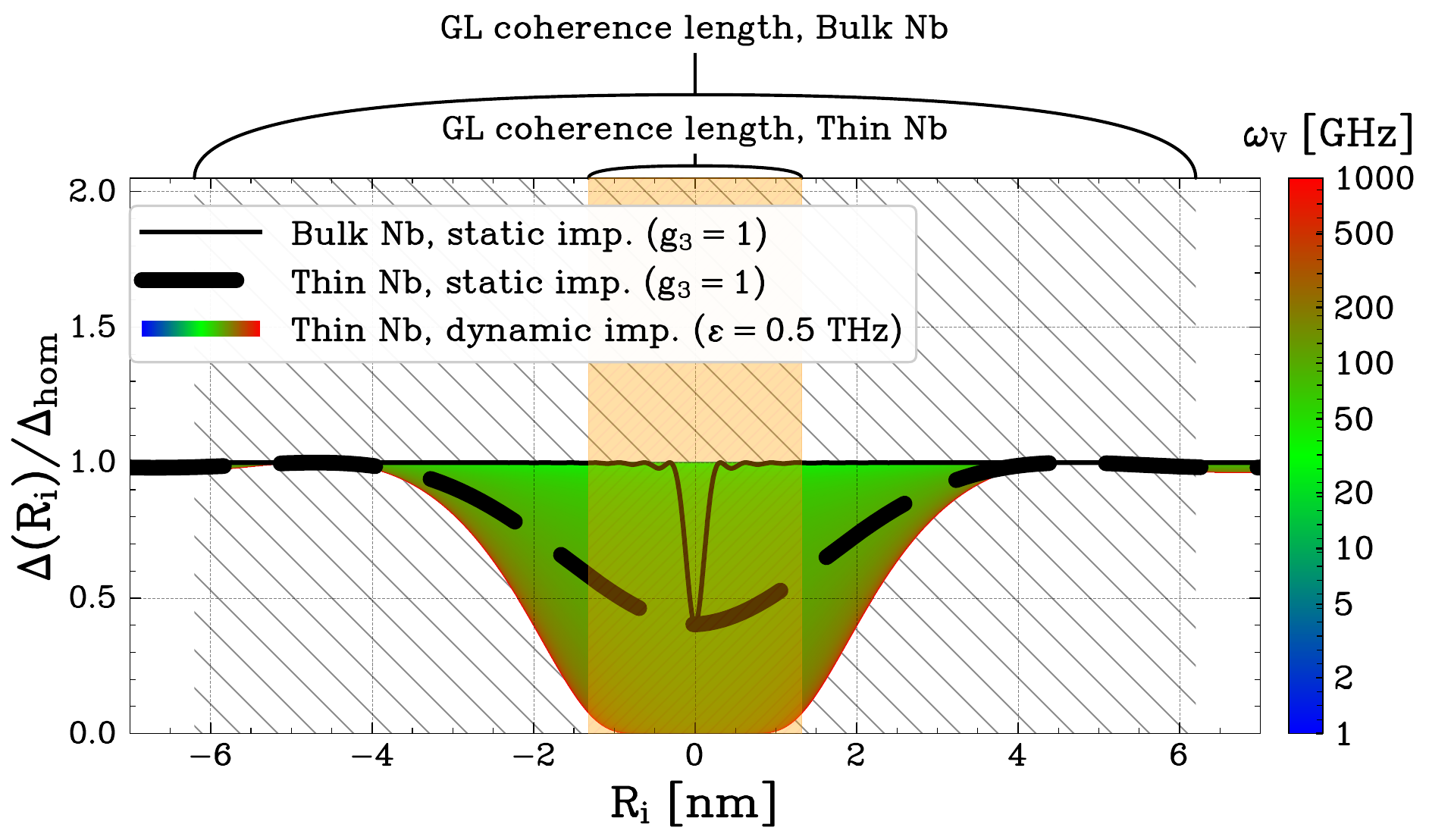}%{V_inelastic.pdf}
\caption{\small The local BCS gap $\Delta({\bf R}_i)$ normalized by the homogeneous BCS gap $\Delta_{\textrm{hom}}$, with an impurity placed at ${\bf R}_i=0$ and the material taken to be Nb. %We take $g_i=\pi N(0)V_i$, with $V_i$ taken to be the impurity potential and $i=0,3$ depending on whether we are talking about a dynamical impurity or a static non-magnetic impurity, respectively. 
%The electron mean free path is taken to be $\ell\sim 50$ nm. 
For the case of a static impurity with strength $g_3=\pi N(0)V_3$ (solid black line), the real space modulation of the gap is significantly smaller than the Ginzburg-Landau (GL) coherence length in bulk Nb (left-hashed background)~\cite{Ilin2014May,Quarterman2020Jul}.
%We take $\xi_{GL}^{\textrm{Nb, bulk}}\approx 12$ nm, as found experimentally for Nb films with thickness $d$ larger than (or on the order of) the coherence length~\cite{Ilin2014May,Quarterman2020Jul}. 
%In ultra-thin film Nb, the electron carrier density is expected to be reduced by around three orders of magnitude as a consequence of dimensional confinement~\cite{Du2004Apr,Ummarino2024Nov}, and thus 
For thin-film Nb, the GL coherence length (orange shaded region) is expected to be reduced due to dimensional confinement~\cite{Du2004Apr,Travaglino2023Jan,Ummarino2024Nov,Zaccone2025May}, and likewise 
%and the ~\cite{Ilin2014May,Quarterman2020Jul,Du2004Apr,Travaglino2023Jan,Ummarino2024Nov,Zaccone2025May}
%and the electron $k_F$ are reduced. From the latter, 
the local modulation of the BCS gap increases (black dashed) (the latter due to a reduction in $k_F$). For the dynamical impurity 
%various vibron frequencies $\omega_V$ are considered 
(multicolor), $\omega_V\approx 100$ GHz results in severe suppression of the gap on the order of the thin film coherence length. 
%the coherence length is expected to shrink as the electron carrier density is reduced by three orders of magnitude as a consequence of dimensional confinement~\cite{Du2004Apr,Ummarino2024Nov}. For the dynamical impurity at various $\omega_V$
\label{fig:gap}}
\end{figure}

To obtain an equation for the subgap bound state energy, we must solve for the poles of the T-matrix. We can do this by assuming the limit of elastic scattering, in which case $\mathcal{T}(i\omega_n,\,i\omega_m;\,{\bf r}_i)=\delta_{nm}\mathcal{T}(i\omega_n;\,{\bf r}_i)$, and Eqn.~\ref{TMatrix} can be easily solved to yield a closed form for the $T$-matrix at some given frequency and real-space position from the impurity. 
%For a physically realistic syst
%[Talk about scattering solution from T-matrix for vibron:]
% \begin{align}
%     \mathcal{T}(i\omega_n;\,{\bf r}_i)=\bigg[-\dfrac{N(0)}{\lambda_V(\omega_V,\,\epsilon)}\tau_0-\mathcal{G}_{\textrm{hom}}(i\omega_n;\,{\bf r}_i)\bigg]^{-1}\tau_0
% \end{align}
The subgap bound state energies are the poles of $\mathcal{T}(i\omega_n;\,{\bf r}_i)$, which correspond to those frequencies where the T-matrix becomes non-invertible; i.e., when $\det[\mathcal{T}^{-1}(i\omega_n;\,{\bf r}_i)]=0$. The full solution for these poles, which we denote as $\omega^*$, are given in Eqn.~\ref{Eqn1}. We immediately see that, as $\omega_V\rightarrow 0$, the poles are confined to the BCS gap $\omega^*=\pm \Delta_0$. %As such, the poles approach $\Delta_0$ in the static limit. 
As we increase the vibron frequency, the poles move away from the gap and cross near zero-energy, as illustrated in Fig.~\ref{fig:boundstatepoles}. 

We can check the validity of Eqn.~\eqref{Eqn1} by considering the effects of the off-shell contributions to the $T$-matrix, thereby going beyond the elastic scattering approximation. This can be done by rewriting the $T$-matrix as a matrix in the Matsubara frequencies ($n,\,m$), which we denote in bold. The self-consistent Lippmann–Schwinger equation can then be written as the following closed equation for the $T$-matrix ${\bf T}$~\cite{Supp}:
\begin{subequations}
\begin{gather}
    {\bf T}={\bf T}_+\otimes \tau_0+{\bf T}_-\otimes \tau_1,\\
    {\bf T}_\pm=\dfrac{1}{2}\left({\bf A}_-^{-1}\pm {\bf A}_+^{-1}\right){\bf V},\\
    {\bf A}_\pm={\bf 1}-{\bf V}\left({\bf G}_{\textrm{hom}}\pm {\bf F}_{\textrm{hom}}\right).
\end{gather}
\end{subequations}
\noindent In the above, ${\bf T}_+$ (${\bf T}_-$) contains information about the effects of scattering in the normal (anomalous) channels. The pole emerges as a Gaussian distortion of $T(i\omega_n,\,i\omega_m)$ in the imaginary frequency plane along the $n=m$ diagonal, and disappears within the same $\omega_V$ range as predicted in Fig.~\ref{fig:boundstatepoles} on the real frequency axis~\cite{Supp}. In Fig.~\ref{fig:V}, we plot the normalized local potential on the Matsubara frequency plane. For $\omega_V$ up to $100-300$ GHz, the potential becomes diagonally dominant, and thus the $T$-matrix will be well approximated by elastic scattering events for vibrons in the GHz regime. 

As the elastic limit accurately approximates the $T$-matrix pole, Eqn.~\eqref{Eqn1} is a good estimate of the subgap bound-state frequency $\omega^*$ in a real material with sub-THz TLS. Table~\ref{tab:Materials} gives worst-case (i.e., lowest expected) estimates for $\omega^*$ assuming $\omega_V=15$ GHz, with data taken from Refs.~\cite{Beck2011Oct,Noat2013Jul,Babu2019Mar,Butler1977Oct,Arnold1980Aug,Carbotte1990Oct,Leung1976,Carbotte1990Oct}. Our results suggest that materials such as In and Sn exhibit substantially lower $\omega^*$ than materials characterized by stronger interatomic bonds, such as NbN.
% As the elastic approximation should give a good estimate for the $T$-matrix pole, Eqn.~\eqref{Eqn1} is a good approximation for the subgap bound state frequency $\omega^*$ in a real material for sub-THz TLSs. In Table~\ref{tab:Materials}, we list estimates for the "worst-case" (i.e., lowest expected) $\omega^*$ for $\omega_V=15$ GHz in several materials. We notice an interesting trend, wherein the lowest expected $\omega^*$ is much lower for materials with weaker interatomic bonds. As such, subgap bound states are expected to be much more prevalent in malleable metals like In and Sn than materials with strong chemical bonds, like NbN.   

% \begin{figure}[t]
% \hspace{-8.75mm}\includegraphics[width=1.05\columnwidth]{T_matrix.png}
% \caption{\small test.
% \label{fig:Tmatrix}}
% \end{figure}

{\it Spectroscopic signatures of dynamical impurities--} Local defects can be probed using local spectroscopy, and often this requires knowledge of the local density of states (LDOS)~\cite{Hess1989Jan,Pan2000Feb,Martin2002Feb,Balatsky2006May,Ji2008Jun}. In a BCS superconductor with a single impurity localized in real space, the total LDOS is given by $N(\omega;\,{\bf R}_i)=N_{\textrm{hom}}(\omega)+\delta N(\omega;\,{\bf R}_i)$, where the homogeneous LDOS is given by $N_{\textrm{hom}}(\omega)=-(1/\pi)\Im[\mathcal{G}_{\hom}(\omega)]_{11} = N_0\omega/{(\omega^2 - \Delta_{\textrm{hom}}^2)^{1/2}}$ (with $N_0$ being the homogeneous normal state DOS), and the modification at a distance ${\bf R}_i={\bf r}-{\bf r}_i$ from the impurity at site ${\bf r}_i$ is given by $
    \delta N(\omega;{\bf R}_i)=-(1/\pi)\Im\left[\mathcal{G}_{\textrm{hom}}(\omega;\,{\bf R}_i)\mathcal{T}(\omega)\mathcal{G}_{\textrm{hom}}(\omega,\,-{\bf R}_i)\right]_{11}
$. Note that the $T$-matrix is calculated on the impurity site; i.e., ${\bf R}_i=0$, and thus we omit the spatial dependence.

\begin{figure*}[t]
\centering

\begin{subfigure}[t]{0.32\textwidth}
\centering
\includegraphics[width=\linewidth]{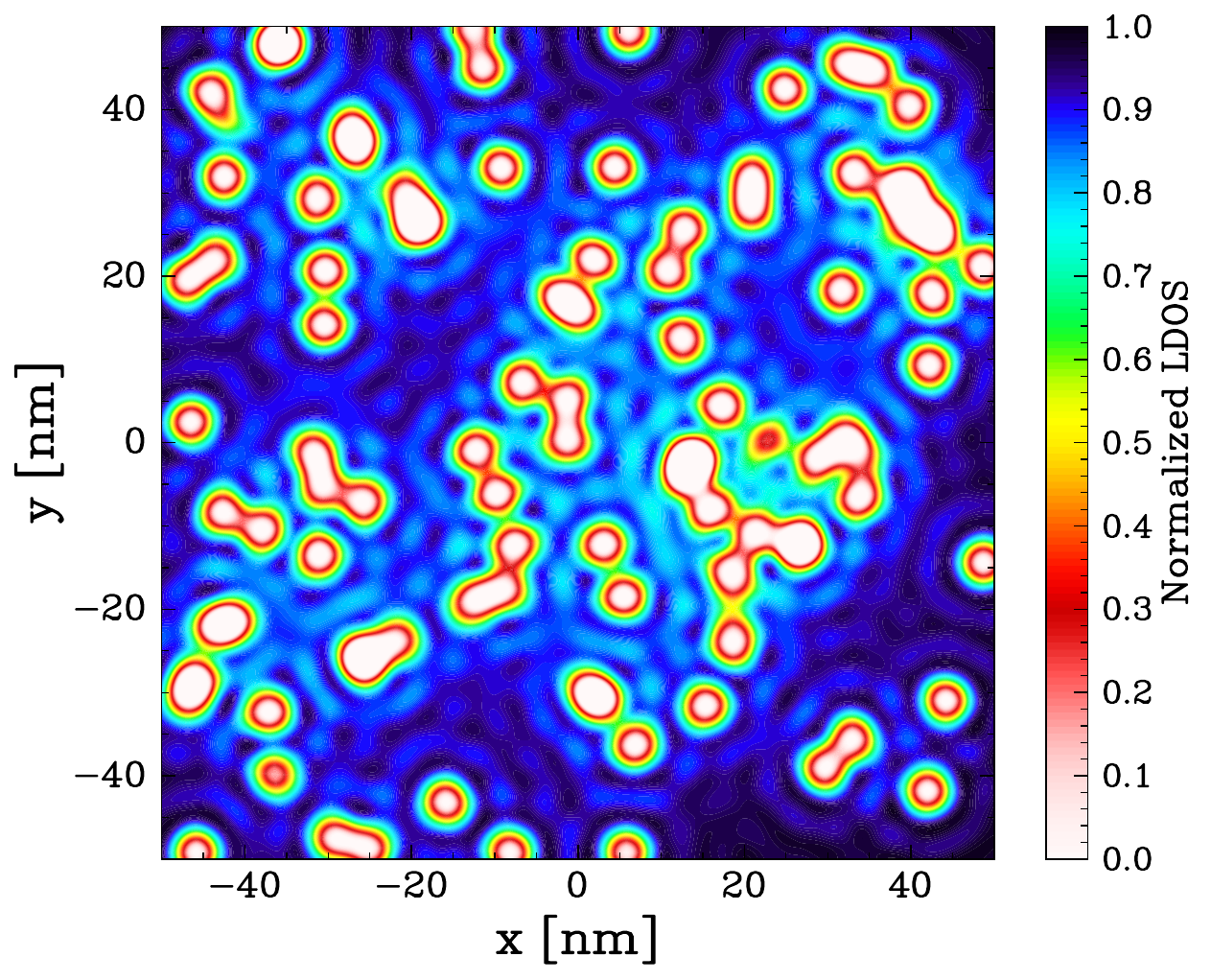}
\caption{LDOS with dynamical impurities, assuming thin film Nb.}
\end{subfigure}
\hfill
\begin{subfigure}[t]{0.32\textwidth}
\centering
\includegraphics[width=\linewidth]{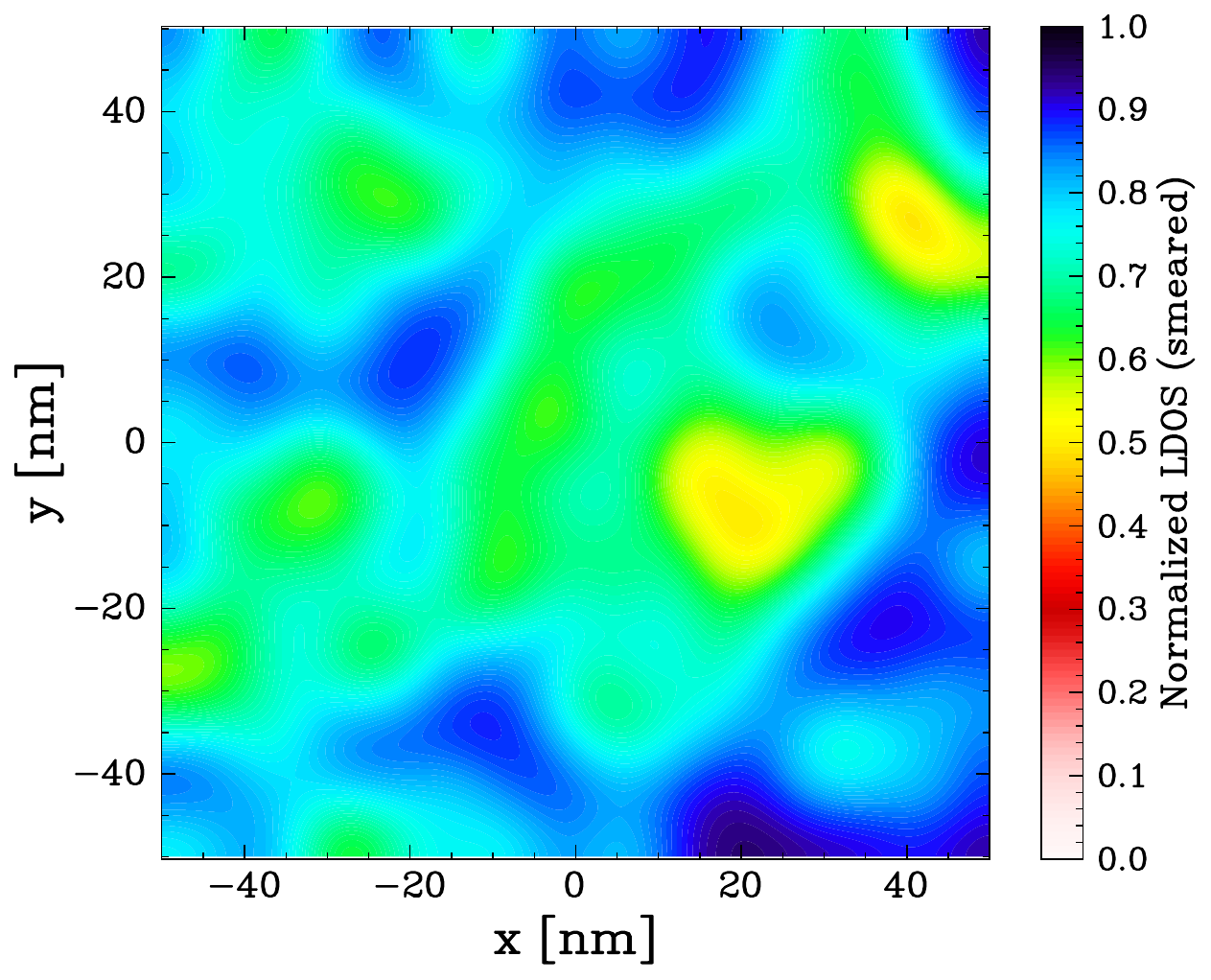}
\caption{Data from (a) "smeared" according to coherence length.}
\end{subfigure}
\hfill
\begin{subfigure}[t]{0.32\textwidth}
\centering
\includegraphics[width=\linewidth]{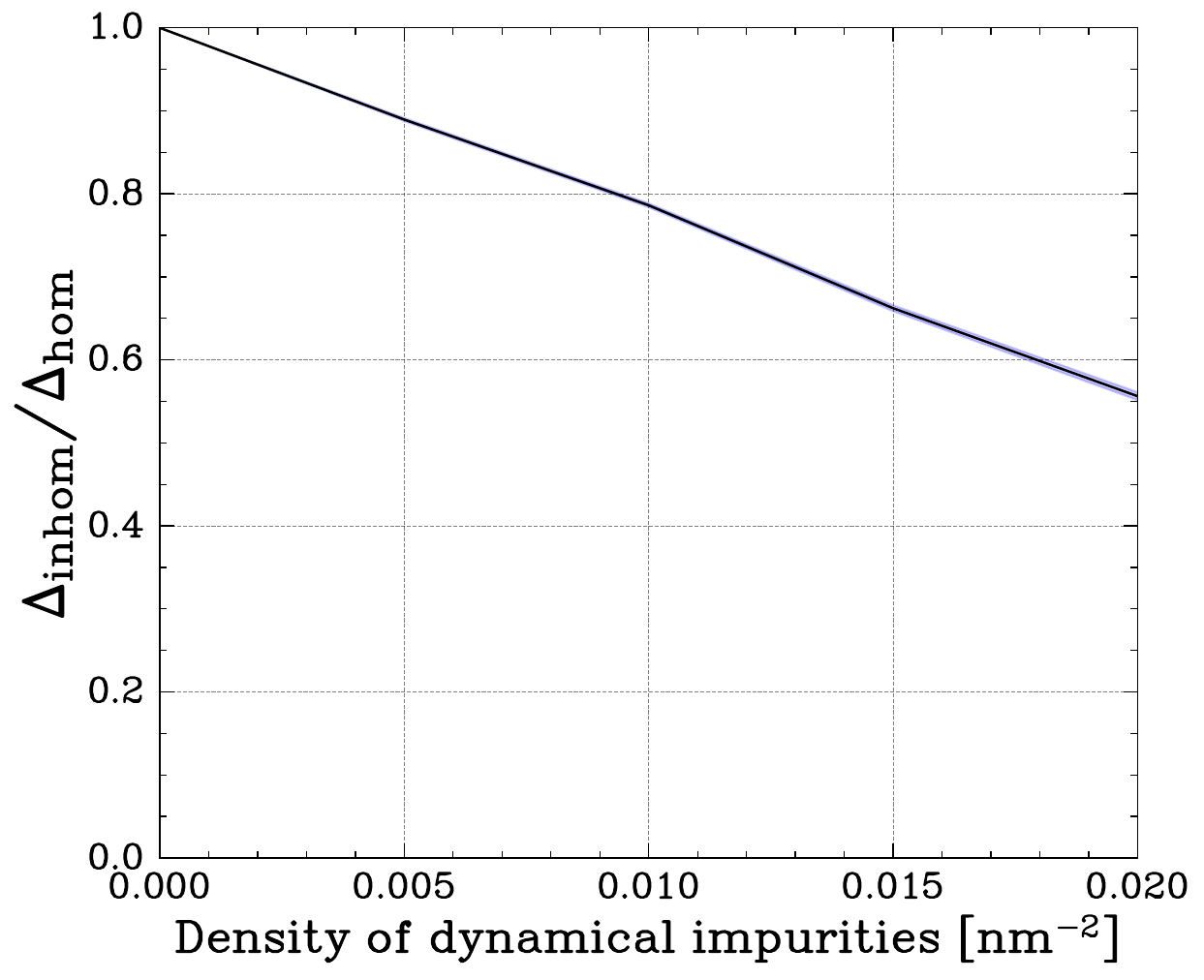}
\caption{Suppression of the average BCS gap with increasing vibron density.}
\end{subfigure}

\caption{\small Collective effects resulting from localized dynamical impurities, with thin-film Nb taken as an example. Parameters are taken from Refs.~\cite{Ilin2014May,Quarterman2020Jul} (same as used in Fig. 5). (a) The normalized local density of states (LDOS) in a 100 nm $\times$ 100 nm square.
%, normalized so that without the impurity the LDOS is unity.
Vibrons (or TLSs) with uniformly random frequencies $\omega_V\in [1,1000]$ GHz and spectral widths $\epsilon\in [500,1000]$ GHz are placed in a random location on the grid. We take the density of vibrons as $n_V=0.01$ vibrons per $nm^{-2}$ ($n_V=10$ $\mu m^{-2}$ GHz$^{-1}$ for our frequency range). This number sits between typical experimental values for the TLS densities in Al/AlO$_x$/Al and Nb/AlO$_x$/Nb junctions~\cite{W.Anderson1972Jan,Wang2025Feb} ($0.4-4.4$ GHz$^{-1}$ $\mu m^{-2}$) and the max possible TLS density on the surface of Nb$_2$O$_5$~\cite{Wang2025Feb} ($20$ GHz$^{-2}$ $\mu m^{-2}$). 
%Strong suppression of the LDOS appears at many of the TLS locations. 
%if we assume that there is a relatively equal distribution of vibrons in the given frequency range.
(b) The same data as in (a) with a Gaussian filter applied to account for the thin-film Nb coherence length ($\xi\sim 2.4$ nm; see Fig.~\ref{fig:gap}). This subfigure thus mimics what a STM would locally measure in a TLS-rich thin film of Nb. (c) Spatially-averaged BCS gap over a $100$ nm $\times$ $100$ nm sample of Nb as a function of vibron density $n_V$, averaged over multiple realization of the filtered LDOS given in (b). At high vibron densities, the averaged BCS gap is suppressed up to $55\%$ of it's homogeneous value.
}
\label{fig:Collective}
\end{figure*}

%LDOS, ldos_avg = contour_plot(kF=.65, l=50, g3=0, density=1, xlim=100/2, ylim=100/2, npoints=500, f=1.001, seed=None)

%always l = 50

%The full derivation for the LDOS is given in the Supplemental Material. 
A general form for the total LDOS is given by~\cite{Supp}
\begin{align}
    \dfrac{N(\omega;\,{\bf R}_i)}{N_{\textrm{hom}}(\omega)}=1-\mathcal{P}^2({\bf R}_i)L_i(\overline{\omega},\,g_i).
\end{align}
In the above, $\overline{\omega}\equiv \omega/\Delta_{\textrm{hom}}$; $\mathcal{P}({\bf R}_i)$ is a spatial modulation parameter dictated by the form of the real-space Green's function (with $\mathcal{P}({\bf R}_i=0)=1$ at the impurity site); $L_\alpha(\omega/\Delta,\,g_\alpha)=[\mathcal{G}_{\textrm{hom}}(\omega)\mathcal{T}(\omega)\mathcal{G}_{\textrm{hom}}(\omega)]_{11}$ is a position-independent term that takes into account the Nambu structure and frequency dependence of the $T$-matrix; and $g_\alpha=\pi N(0)V_\alpha$ is a dimensionless interaction corresponding to the $\tau_\alpha$ impurity. 

For the $\tau_3$ impurity, we reproduce $L_3(\overline{\omega},\,g_3)=L_3(g_3)=g_3^2/(g_3^2+1)$~\cite{Zarea2023Oct}.
%, which is generally a known result. 
%For the $\tau_3$ impurity, the LDOS is characterized by a frequency-independent suppression. 
In the limit $g_3\rightarrow \infty$, the LDOS vanishes at the impurity site for all frequency. However, Anderson’s theorem~\cite{Anderson1959Sep} remains applicable due to the spatial modulation $\mathcal{P}^2({\bf R}_i)$ suppressing the LDOS modulation beyond $\sim 0.5$ nm (well below the Ginzburg-Landau (GL) coherence length for most conventional superconductors). Note that superconducting thin films~\cite{Travaglino2023Jan,Zaccone2025May} may be characterized by a reduced carrier density~\cite{Du2004Apr,Ummarino2024Nov}, thereby lowering the GL coherence length such that the Cooper pair size becomes comparable to the LDOS modulation in certain materials. Nevertheless, strong screening in conventional metals keeps $g_3$ small, with self-consistent estimates for Nb yielding $g_3\lessapprox 1$~\cite{Flatte1997Nov} (see Fig.~\ref{fig:gap}).

For the dynamical $\tau_0$ impurity, we find that~\cite{Supp}
%we find that $L_0(\omega/\Delta_{\textrm{hom}},\,g_0)$ now goes as
%case of the $\tau_3$ and $\tau_0$ impurities, the parameter becomes the following
%  \begin{figure}[t]
% \hspace{-5.75mm}\includegraphics[width=1.05\columnwidth]{waterfall.pdf}
% \caption{\small test.
% \label{fig:lambda}}
% \end{figure}

\begin{align}
&L_0\left(\overline{\omega},\,g_0\right)=\dfrac{g_0^2}{g_0^2+1}\cdot \dfrac{\overline{\omega}^2-1+4/(g_0^2+1)
}{
\overline{\omega}^2-1
+4g_0^2/(g_0^2+1)^2
}.
\end{align}

\noindent Two key differences arise for the dynamical impurity compared to the static $\tau_3$ impurity. First, note that the function $L_0(\omega/\Delta, g_0)$ exhibits strong frequency dependence, with $\lim_{\omega/\Delta_{\mathrm{hom}} \rightarrow 1} L_0(\omega/\Delta_{\textrm{hom}},\,g_0) = 1$. This implies that, for frequencies just above the homogeneous BCS gap, the LDOS is strongly suppressed independent of $g_0$ (see Fig.~\ref{fig:waterfall}). Our result is true for {any} local dynamical potential $g_0$, {\it regardless of the potential's strength $g_0$}. A similar depletion of energy states above the superconducting gap has been predicted to emerge via Andreev levels~\cite{Levchenko2008May}. Second, since $g_0 \equiv \pi \lambda(\omega_V, \epsilon)\sim 1-10$ for experimentally realistic parameters of $\omega_V$ and $\epsilon$~\cite{Supp}, the suppression of the LDOS from a generic $\tau_0$ impurity is expected to be significantly more pronounced than in the static case. Fig.~\ref{fig:Collective} illustrates this for a model of a $100\,\textrm{nm}\times 100\,\textrm{nm}$ thin-film Nb sample containing randomly distributed vibrons~\cite{Wang2025Feb,Ilin2014May,Quarterman2020Jul}. Experimentally verifiable inhomogeneity of the LDOS persists on length scales of tens of nm, producing a noticeable suppression of the spatially averaged BCS gap.

{\it Conclusions}-- Non-magnetic disorder is ubiquitous in real-world materials. For static impurities, any effects on the local density of states of conventional superconductors will typically be on length scales much smaller than the Ginzburg-Landau coherence length. We have shown that this is not the case for dynamical defects (i.e. vibrons). We have shown that a single dynamical defect with frequency $\omega_V$ in the sub-THz regime will generically result in i) subgap bound states with frequencies $\pm \omega^*$, and ii) prominent LDOS modulation above the gap that should prove detectable in certain thin film superconductors and at high enough impurity densities. Using a simple model of a localized dynamical impurity~\cite{Hewson2001Dec,Slager2015Aug,Denisov2024Jul,He2025Jun,Al-Eryani2026Apr}, we suggest that vibrons can produce experimentally accessible signatures in the thin-film limit of conventional superconductors, thus demonstrating that generic inhomogeneity of the local energy landscape can be a significant source of YSR-like bound states.

{\it Acknowledgements}-- The author would like to thank Alexander Balatsky and Rufus Boyack for thorough readings of the paper. In addition, JTH is very thankful for useful feedback from Graham Kells, Gabriele Naselli, Pavel Volkov, and Evan Wilson, in addition to the hospitality of the Roost Maynooth. Prior to April 30th 2026, this work was supported by Novonordisk NQCP Novo Nordisk Foundation, Grant number NNF22SA0081175. As of May 1st 2026, JTH has been funded by a Pathway Grant from Taighde \'Eireann--Research Ireland, and is employed by Maynooth University. Specifically, this publication has emanated from research conducted with the financial support of Taighde \'Eireann--Research Ireland, Grant Number 24/PATH-S/12630. For the purpose of Open Access, the author has applied a CC BY public copyright licence to any Author Accepted Manuscript (AAM) version arising from this submission. The AAM will be available for free on the arXiv (a free distribution service and an open-access archive) upon completion of peer review. The original manuscript will also be available for free on the arXiv prior to completion of peer review.

\bibliography{main}{}

\end{document}

\title{Supplemental Material: \\ Subgap Bound States 
%and Spectroscopic Signatures\\  
from Dynamical Impurities}

\author{Joshuah T. Heath}
\email{joshuah.heath@mu.ie}

\affiliation{Department of Physics, Science Building, Maynooth University, Maynooth, Co. Kildare W23 F2H6, Ireland}
\affiliation{Hamilton Institute, Eolas Building, Maynooth University, Maynooth, Co. Kildare W23 A3HY, Ireland}
%\affiliation{Trinity Quantum Alliance, Unit 16, Trinity Technology and Enterprise Centre, Pearse Street, Dublin 2, D02 YN67, Ireland}
\affiliation{Nordita, Stockholm University and KTH Royal Institute of Technology, Hannes Alfvéns väg 12, SE-106 91 Stockholm, Sweden}
\affiliation{Department of Physics, University of Connecticut, Storrs, Connecticut 06269, USA}

% \author{Alexander V. Balatsky}
% \email{balatsky@kth.se}

% \affiliation{Nordita, Stockholm University and KTH Royal Institute of Technology, Hannes Alfvéns väg 12, SE-106 91 Stockholm, Sweden}
% \affiliation{Department of Physics, University of Connecticut, Storrs, Connecticut 06269, USA}
% \affiliation{Institute for Materials Science, Storrs, University of Connecticut, 06269, USA}

\date{\today}

% \begin{abstract}
% \noindent We argue that a single non-magnetic dynamical impurity 
% (i.e., a low-frequency "vibron" mode localized in real space) 
% results in a subgap bound state in a s-wave superconductor. 
% %We show this explicitly by calculating the poles of the $T$-matrix induced by a localized "vibron" excitation. 
% A closed-form solution for the bound state energy is found as a function of the vibron frequency in the elastic scattering limit, with the salient features of the bound state energies unaffected by inelastic scattering processes in the sub-THz regime. Such impurities result in real space regions characterized by sharp spectroscopic features outside the gap peak and a low-energy density of states.
% %real-space characterized by a low-energy density of states.
% %at high impurity concentration.
% %, and suggest that dynamical impurities may serve as a      
% \end{abstract}

\pacs{1}

\begin{abstract}
\noindent 
%\blindtext
\end{abstract}

\pacs{1}

\maketitle

\renewcommand{\theequation}{S.\arabic{equation}}

\tableofcontents

\section{Subgap bound states: the $T$-matrix for $|\omega|<\Delta$}

\subsection{The superconducting Green's function in real space}

The overarching goal of the paper is twofold: i) to calculate the frequency of subgap poles from a dynamical impurity, and ii) to calculate the change in the local density of states (LDOS) resulting from these impurities. To find these quantities, we have to understand the form of the $T$-matrix below ($|\omega|<\Delta$) and above ($|\omega|>\Delta$) the BCS gap. 

First, we define the homogeneous superconducting Green's function in the Nambu basis~\cite{RickayzenBook}:

\begin{align}
    \mathcal{G}_{\textrm{hom}}(\omega,\,{\bf k})&=\dfrac{\omega \tau_0+\xi_k\tau_3 +\Delta_k \tau_1}{\omega^2-\xi_k^2-\Delta_k^2}\notag\\
%    &=\dfrac{\omega \tau_0 +\xi_k \tau_3}{\omega^2-\xi_k^2-\Delta_k^2}+\dfrac{\Delta_k\tau_1}{\omega^2-\xi_k^2-\Delta_k^2}\notag\\
    &\equiv G_0(\omega,\,{\bf k})\tau_0+G_3(\omega,\,{\bf k})\tau_3+F_1(\omega,\,k)\tau_1.
\end{align}
\noindent where $\xi_k =\epsilon_k-\epsilon_F$. In the above, we have split up the superconducting Green's function into three components: a term which goes as the frequency $\omega$ ($\tau_0$), a term which goes linearly with the dispersion $\xi_k$ ($\tau_3$), and a term which is proportional to the gap $\Delta_k$ ($\tau_1$). For the purposes of our work, we restrict ourselves to s-wave superconductors, and thus in $k$-space we write the 2-component Nambu spinor as $\Psi_k=\begin{pmatrix}c_{k\uparrow}&c_{-k\downarrow}^\dagger \end{pmatrix}^T$, or equivalently $\Psi_k^\dagger=\begin{pmatrix}c_{k\uparrow}^\dagger &c_{-k\downarrow} \end{pmatrix}$.

For the purposes of impurity scattering, we are interested in the effects of a single impurity (as opposed to an average), and thus we want to solve for the homogeneous superconducting Green's function in real space. This is done below, where we take ${\bf R}_i={\bf r}-{\bf r}_i$ as the real space distance from the impurity at site ${\bf r}_i$:

\begin{align}
    \mathcal{G}_0(\omega,\,{\bf R}_i)&=\sum_k e^{i{\bf k}\cdot {\bf R}_i}\dfrac{\omega}{(\omega+i\delta)^2-\xi_k^2-\Delta^2}\notag\\
    &\rightarrow \int \dfrac{d^3 k}{(2\pi)^3}e^{i{\bf k}\cdot {\bf R}_i}\dfrac{\omega}{(\omega+i\delta)^2-\xi_k^2-\Delta^2}\notag\\
%    &=\int_0^\infty \dfrac{k^2 dk}{(2\pi)^3}\dfrac{\omega}{(\omega+i\delta)^2-\xi_k^2-\Delta^2}\int d\Omega e^{i{\bf k}\cdot {\bf R}_i\cos\theta}\notag\\
    &= \int_0^\infty \dfrac{k^2dk}{2\pi^2}\dfrac{\omega\sin(kR_i)/(kR_i)}{(\omega+i\delta)^2-\xi_k^2-\Delta^2}.
\end{align}
Linearizing near the Fermi surface we have $k\approx k_F$, and thus $\xi_k=v_F(k-k_F)\rightarrow k^2dk\approx k_F^2d\xi_k/v_F,\qquad k=k_F+\xi_k/v_F$. Noting that $k_F^2/v_F=(2\pi^2)N(0)$, the above simplifies to

\begin{align}
    \int_0^\infty \dfrac{k^2dk}{2\pi^2}\dfrac{\omega\sin(kR_i)/(kR_i)}{(\omega+i\delta)^2-\xi_k^2-\Delta^2}
    %&=\dfrac{k_F^2}{2\pi^2v_F k_FR_i}\int_{-\infty}^{\infty}d\xi_k \dfrac{\omega\sin[(k_F+\xi_k/v_F)R_i]}{(\omega+i\delta)^2-\xi_k^2-\Delta^2}\notag\\
    &=\dfrac{N(0)}{k_FR_i}\int_{-\infty}^{\infty}d\xi_k\dfrac{\omega \bigg[\sin(k_F R_i)\cos(\xi_kR_i/v_F)+\cos(k_F R_i)\sin(\xi_kR_i/v_F)\bigg]}{(\omega+i\delta)^2-\xi_k^2-\Delta^2}.
\end{align}

If we take $\xi_k R_i/v_F=(k-k_F)R_i\approx 0$, and thus the above becomes 

\begin{align}
    G_0(\omega,\,{\bf R}_i)&=N(0)\dfrac{\sin(k_F R_i)}{k_FR_i}\int d\xi_k \dfrac{\omega}{(\omega+i\delta)^2-\xi_k^2-\Delta^2}\notag\\
    &=-\pi N(0)\textrm{sgn}(\omega)\dfrac{\sin(k_F R_i)}{k_F R_i}\dfrac{\omega}{\sqrt{\Delta^2-\omega^2}}.
\end{align}

In the limit of ${\bf R}_i\rightarrow 0$, the above simplifies to 
\begin{align}
    G_0(\omega;\,0)&=-i\pi N(0)\dfrac{\omega }{\sqrt{\omega^2-\Delta^2}}\notag\\
    &=-\pi N(0)\textrm{sgn}(\omega)\dfrac{\omega}{\sqrt{\Delta^2-\omega^2}}.
\end{align}
\noindent In the above, we have carefully used the fact that $\sqrt{\Delta^2-\omega^2}=-i\textrm{sgn}(\omega)\sqrt{\omega^2-\Delta^2}$. The sign may be identified by carefully separating the imaginary component:
\begin{align}
    \sqrt{\Delta^2-(\omega+i\delta)^2}&\approx\sqrt{-(\omega^2-\Delta^2)-2i\omega \delta }\notag\\
    &\equiv \sqrt{-x\mp i\delta_\omega}\notag\\
    &\approx \sqrt{x}e^{\mp i\pi/2}\notag\\
    &=\mp i\sqrt{x}\notag\\
    &=-\textrm{sgn}(\omega)i\sqrt{\omega^2-\Delta^2}.
\end{align}
Note that the above gives the form of $G_0(\omega;\,{\bf R}_i)$, which is the frequency-dependent component of the superconducting Green's functions. Without loss of generality, we similarly find that
\begin{align}
    G_3(\omega;\,{\bf R}_i)=0,
\end{align}

\begin{align}
    F_1(\omega;\,{\bf R}_i)=-\pi N(0)\textrm{sgn}(\omega)\dfrac{\sin(k_F R_i)}{k_F R_i}\dfrac{\Delta}{\sqrt{\Delta^2-\omega^2}}.
\end{align}
In addition to the above, we can include a small decay term from the mean free path. The mean free path of the electron results in $k\rightarrow k+i\Gamma/v_F$, where $\Gamma=1/(2\tau)=v_F/(2\ell)$.  This results in a damping term of $e^{-R_i/(2\ell)}$ being attached to the above, and thus for a finite distance away from the impurity all propagators are modified by the function
\begin{align}
    \mathcal{P}({\bf R}_i)=e^{-R_i/2\ell}\dfrac{\sin(k_F R_i)}{k_F R_i}.
\end{align}

\noindent As introduced in the main text.

\begin{figure*}
\centering

\[
\begin{tikzpicture}[baseline=-2pt]
  \begin{feynman}
    \vertex (i);
    \vertex [right=2cm of i] (o);
    \diagram*{
      (i) --[
        double,
        double distance=0.5ex,
        thick,
        with arrow=0.5,
        arrow size=0.3em,
        postaction={
          decorate,
          decoration={
            markings,
            mark=at position 0.5 with {
              \node[below=6pt]{\smash{\(n\)}};
            }
          }
        }
      ] (o)
    };
  \end{feynman}
\end{tikzpicture}
~=~
\begin{tikzpicture}[baseline=-2pt]
  \begin{feynman}
    \vertex (i);
    \vertex [right=2cm of i] (o);
    \diagram*{
      (i) --[
        fermion,
        postaction={
          decorate,
          decoration={
            markings,
            mark=at position 0.5 with {
              \node[below=6pt]{\smash{\(n\)}};
            }
          }
        }
      ] (o)
    };
  \end{feynman}
\end{tikzpicture}
~+~
\begin{tikzpicture}[baseline=-2pt]
  \begin{feynman}
    \vertex (i);
    \node[right=2cm of i,draw,fill=white,circle,inner sep=2pt] (v){$\Sigma$};
    \vertex [right=2cm of v] (o);
    \diagram*{
      (i) --[
        fermion,
        postaction={
          decorate,
          decoration={
            markings,
            mark=at position 0.5 with {
              \node[below=6pt]{\smash{\(n\)}};
            }
          }
        }
      ] (v),
      (v) --[
        double,
        double distance=0.5ex,
        thick,
        with arrow=0.5,
        arrow size=0.3em,
        postaction={
          decorate,
          decoration={
            markings,
            mark=at position 0.5 with {
              \node[below=6pt]{\smash{\(n\)}};
            }
          }
        }
      ] (o)
    };
  \end{feynman}
  \node[below=6pt of current bounding box.south]
  {\text{\hspace{-59mm}(a) Local self-energy $\Sigma(\omega_n)$ for some interacting fermionic system.}};
\end{tikzpicture}
\]

\vspace{5mm}

\[
\hspace{12mm}
\begin{tikzpicture}[baseline=-1pt]
  \begin{feynman}
    \vertex (i);
    \vertex [right=2cm of i] (o);
    \diagram*{
      (i) --[
        double,
        double distance=1.2ex,
        line width=1pt,
        with arrow=0.18,
        with arrow=0.82,
        arrow size=0.4em,
        postaction={
          decorate,
          decoration={
            markings,
            mark=at position 0.5 with {
              \draw[line width=1.2pt]
                (-2.9ex,0.72ex)
                .. controls (0.9ex,0.25ex)
                         and (-0.9ex,-0.25ex) ..
                (2.9ex,-0.72ex);
            }
          }
        },
        postaction={
          decorate,
          decoration={
            markings,
            mark=at position 0.18 with {
              \node[below=6pt]{\smash{\(n\)}};
            },
            mark=at position 0.82 with {
              \node[below=6pt]{\smash{\(m\)}};
            }
          }
        }
      ] (o)
    };
  \end{feynman}
\end{tikzpicture}
~=~
\begin{tikzpicture}[baseline=-1pt]
  \begin{feynman}
    \vertex (i);
    \vertex [right=2cm of i] (o);
    \diagram*{
      (i) --[
        double,
        double distance=0.5ex,
        thick,
        with arrow=0.5,
        arrow size=0.3em,
        postaction={
          decorate,
          decoration={
            markings,
            mark=at position 0.5 with {
              \node[below=7pt]{\smash{\hspace{3mm}\(n\delta_{nm}\)}};
            }
          }
        }
      ] (o)
    };
  \end{feynman}
\end{tikzpicture}
~+~
\hspace{-10.5mm}
\begin{tikzpicture}[baseline=-1pt]
  \begin{feynman}
    \vertex (i);
    \vertex [right=2cm of i] (v);
    \vertex [right=2cm of v] (o);
    \vertex [above=1.3cm of v] (x);
    \diagram*{
      (i) --[
        double,
        double distance=0.5ex,
        thick,
        with arrow=0.5,
        arrow size=0.3em,
        postaction={
          decorate,
          decoration={
            markings,
            mark=at position 0.5 with {
              \node[below=6pt]{\smash{\(n\)}};
            }
          }
        }
      ] (v),
      (v) --[
        double,
        double distance=0.5ex,
        thick,
        with arrow=0.5,
        arrow size=0.3em,
        postaction={
          decorate,
          decoration={
            markings,
            mark=at position 0.5 with {
              \node[below=6pt]{\smash{\(m\)}};
            }
          }
        }
      ] (o),
      (v) --[dashed] (x),
    };
    \node[circle, fill=black, inner sep=0pt, minimum size=4.5pt] at (v) {};
    \draw[line width=1pt] ($(x)+(-0.16,-0.16)$) -- ($(x)+(0.16,0.16)$);
    \draw[line width=1pt] ($(x)+(-0.16,0.16)$) -- ($(x)+(0.16,-0.16)$);
  \end{feynman}
  \node[below=6pt of current bounding box.south]
  {\text{\hspace{-58mm}\phantom{(b) Local $T$-matrix $\mathcal{T}(\omega_n,\omega_m)$ for translation-symmetry-breaking inelastic scattering}}};
\end{tikzpicture}
\hspace{-10mm}~+~
\begin{tikzpicture}[baseline=-1pt]
  \begin{feynman}
    \vertex (i2);
    \vertex [right=1.6cm of i2] (va);
    \vertex [right=1.6cm of va] (vb);
    \vertex [right=1.6cm of vb] (o2);
    \vertex (xapex) at ($(va)!0.5!(vb)+(0,1.3cm)$);
    \diagram*{
      (i2) --[
        double,
        double distance=0.5ex,
        thick,
        with arrow=0.5,
        arrow size=0.3em,
        postaction={
          decorate,
          decoration={
            markings,
            mark=at position 0.5 with {
              \node[below=6pt]{\smash{\(n\)}};
            }
          }
        }
      ] (va),
      (va) --[
        double,
        double distance=0.5ex,
        thick,
        with arrow=0.5,
        arrow size=0.3em,
        postaction={
          decorate,
          decoration={
            markings,
            mark=at position 0.5 with {
              \node[below=6pt]{\smash{\(\ell\)}};
            }
          }
        }
      ] (vb),
      (vb) --[
        double,
        double distance=0.5ex,
        thick,
        with arrow=0.5,
        arrow size=0.3em,
        postaction={
          decorate,
          decoration={
            markings,
            mark=at position 0.5 with {
              \node[below=6pt]{\smash{\(m\)}};
            }
          }
        }
      ] (o2),
      (va) --[dashed] (xapex),
      (vb) --[dashed] (xapex),
    };
    \node[circle, fill=black, inner sep=0pt, minimum size=4.5pt] at (va) {};
    \node[circle, fill=black, inner sep=0pt, minimum size=4.5pt] at (vb) {};
    \draw[line width=1pt] ($(xapex)+(-0.16,-0.16)$) -- ($(xapex)+(0.16,0.16)$);
    \draw[line width=1pt] ($(xapex)+(-0.16,0.16)$) -- ($(xapex)+(0.16,-0.16)$);
  \end{feynman}
\end{tikzpicture}
~+~{\LARGE \cdots}
\]

\vspace{-20mm}

\[
\hspace{12mm}
\phantom{\begin{tikzpicture}[baseline=-1pt]
  \begin{feynman}
    \vertex (i);
    \vertex [right=2cm of i] (o);
    \diagram*{
      (i) --[
        double,
        double distance=1.2ex,
        line width=1pt,
        with arrow=0.18,
        with arrow=0.82,
        arrow size=0.4em,
        postaction={
          decorate,
          decoration={
            markings,
            mark=at position 0.5 with {
              \draw[line width=1.2pt]
                (-2.9ex,0.72ex)
                .. controls (0.9ex,0.25ex)
                         and (-0.9ex,-0.25ex) ..
                (2.9ex,-0.72ex);
            }
          }
        },
        postaction={
          decorate,
          decoration={
            markings,
            mark=at position 0.18 with {
              \node[below=6pt]{\smash{\(n\)}};
            },
            mark=at position 0.82 with {
              \node[below=6pt]{\smash{\(m\)}};
            }
          }
        }
      ] (o)
    };
  \end{feynman}
\end{tikzpicture}}
\hspace{1.2mm}~=~\hspace{2mm}
\begin{tikzpicture}[baseline=-1pt]
  \begin{feynman}
    \vertex (i);
    \vertex [right=2cm of i] (o);
    \diagram*{
      (i) --[
        double,
        double distance=0.5ex,
        thick,
        with arrow=0.5,
        arrow size=0.3em,
        postaction={
          decorate,
          decoration={
            markings,
            mark=at position 0.5 with {
              \node[below=7pt]{\smash{\hspace{3mm}\(n\delta_{nm}\)}};
            }
          }
        }
      ] (o)
    };
  \end{feynman}
\end{tikzpicture}
\hspace{5mm}~+~
\hspace{-30mm}
\begin{tikzpicture}[baseline=-1pt]
% \hspace{-12mm}%\begin{tikzpicture}[baseline=-1pt]
  \begin{feynman}
    \vertex (i);
    \node[right=2cm of i,draw,fill=white,circle,inner sep=2pt] (v){$\mathcal{T}$};
    \vertex [right=2cm of v] (o);
    \diagram*{
      (i) --[double,double distance=0.5ex,thick,with arrow=0.5,arrow size=0.3em,
             postaction={decorate,decoration={markings,
               mark=at position 0.5 with {\node[below=6pt]{\smash{\(n\)}};}}
             }] (v),
      (v) --[double,double distance=0.5ex,thick,with arrow=0.5,arrow size=0.3em,
             postaction={decorate,decoration={markings,
               mark=at position 0.5 with {\node[below=6pt]{\smash{\(m\)}};}}
             }] (o)
    };
  \end{feynman}
  \node[below=10pt of current bounding box.south]
  {\text{\hspace{-19mm}{\phantom{dfdfdffdfsds dsf}(b) Local $T$-matrix $\mathcal{T}(\omega_n,\omega_m)$ in the presence of an impurity.\phantom{dfjfk sfkjfksdjf } }}};
\end{tikzpicture}
\hspace{-0mm}
\phantom{~+~
\begin{tikzpicture}[baseline=-1pt]
  \begin{feynman}
    \vertex (i2);
    \vertex [right=1.6cm of i2] (va);
    \vertex [right=1.6cm of va] (vb);
    \vertex [right=1.6cm of vb] (o2);
    \vertex (xapex) at ($(va)!0.5!(vb)+(0,1.3cm)$);
    \diagram*{
      (i2) --[
        double,
        double distance=0.5ex,
        thick,
        with arrow=0.5,
        arrow size=0.3em,
        postaction={
          decorate,
          decoration={
            markings,
            mark=at position 0.5 with {
              \node[below=6pt]{\smash{\(n\)}};
            }
          }
        }
      ] (va),
      (va) --[
        double,
        double distance=0.5ex,
        thick,
        with arrow=0.5,
        arrow size=0.3em,
        postaction={
          decorate,
          decoration={
            markings,
            mark=at position 0.5 with {
              \node[below=6pt]{\smash{\(\ell\)}};
            }
          }
        }
      ] (vb),
      (vb) --[
        double,
        double distance=0.5ex,
        thick,
        with arrow=0.5,
        arrow size=0.3em,
        postaction={
          decorate,
          decoration={
            markings,
            mark=at position 0.5 with {
              \node[below=6pt]{\smash{\(m\)}};
            }
          }
        }
      ] (o2),
      (va) --[dashed] (xapex),
      (vb) --[dashed] (xapex),
    };
    \node[circle, fill=black, inner sep=0pt, minimum size=4.5pt] at (va) {};
    \node[circle, fill=black, inner sep=0pt, minimum size=4.5pt] at (vb) {};
    \draw[line width=1pt] ($(xapex)+(-0.16,-0.16)$) -- ($(xapex)+(0.16,0.16)$);
    \draw[line width=1pt] ($(xapex)+(-0.16,0.16)$) -- ($(xapex)+(0.16,-0.16)$);
  \end{feynman}
\end{tikzpicture}}
\phantom{~+~{\LARGE \cdots}}
\]

\caption{\small Diagrammatic formulation of the $T$-matrix formalism. In (a), we show the diagrammatic form of the self-consistent Dyson equation, which is contrasted in (b) where we illustrate the Lippman-Schwinger equation for a generic potential represented by an $X$. The interaction between the Green's function and the impurity is represented by a dashed line. As we consider a purely dynamical very low-frequency impurity (a vibron) in a BCS superconductor, we ignore further vertex corrections for the effective potential.}
\label{fig:Feynman}

\end{figure*}
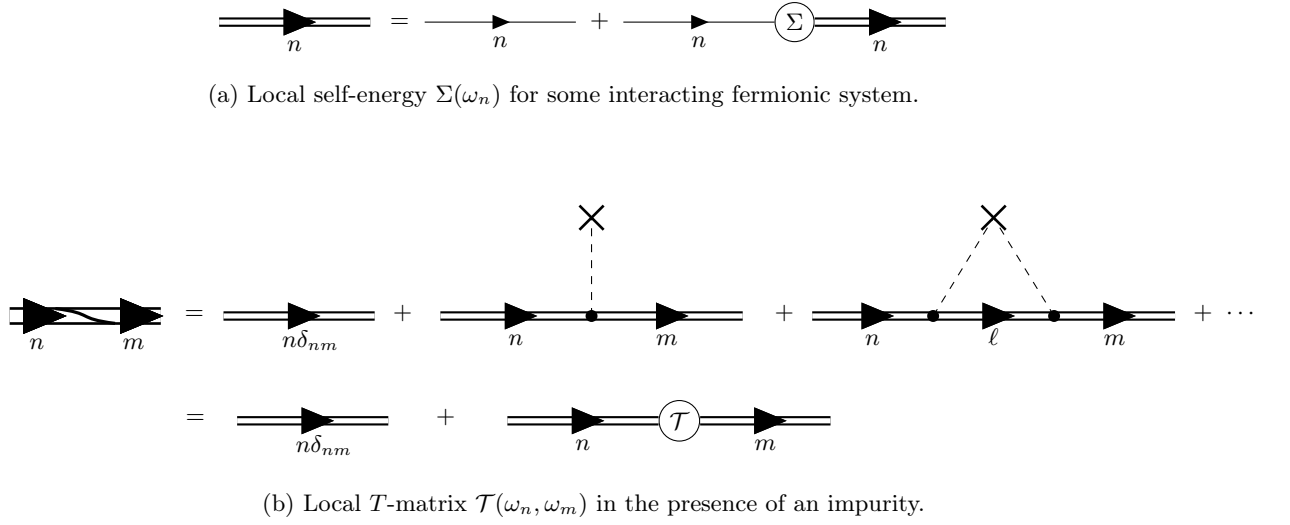

\subsection{Subgap pole structure in the elastic scattering limit}

If we assume scattering from an impurity, then we may write a general form for the inhomogeneous Green's function as a self-consistent equation in terms of the homogeneous and the superconducting $T$-matrix. We will write the inhomogeneous Green's function below, which we'll write in real space in terms of local propagators and make the $r$-dependence explicit where appropriate~\cite{Hirschfeld1986Jul,Hirschfeld1993Aug,Hotta1993Dec,Ziegler1996Apr,Hussey2002Dec,Bruus2004Sep,Balatsky2006May,Bena2016Mar}:

\begin{align}
    \mathcal{G}(\omega_n,\,\,\omega_m;\,{\bf r},\,{\bf r}')=\mathcal{G}_{\textrm{hom}}(\omega_n;\,{\bf r}-{\bf r}')\delta_{n,\,m}+\mathcal{G}_{\textrm{hom}}(\omega_n;\,{\bf r}-{\bf r}_i)\mathcal{T}(\omega_n,\,\omega_m;\,{\bf r}_i)\mathcal{G}_{\textrm{hom}}(\omega_m;\,{\bf r}_i-{\bf r}').
\end{align}
\noindent Note that the $T$-matrix $\mathcal{T}(\omega_n,\,\omega_m;\,{\bf r}_i)$ is calculated on the impurity site, and thus we drop the spatial coordinate. Note that the above is a very general form, as it defines an inelastic scattering of an electron with frequency $\omega_n$ to $\omega_m$ from an impurity at site ${\bf r}={\bf r}_i$. In the elastic scattering limit, $\omega_n\approx \omega_m$, and thus we may write

\begin{align}
    \mathcal{G}(\omega_n;\,{\bf r},\,{\bf r}')=\mathcal{G}_{\textrm{hom}}(\omega_n;\,{\bf r}-{\bf r}')\delta_{n,\,m}+\mathcal{G}_{\textrm{hom}}(\omega_n;\,{\bf r}-{\bf r}_i)\mathcal{T}(\omega_n;\,{\bf r}_i)\mathcal{G}_{\textrm{hom}}(\omega_n,\,{\bf r}_i-{\bf r}').
\end{align}

%In the future, we will continue to use the notation of no ${\bf r}$-dependence to denote calculation on the impurity site.

\noindent In the elastic scattering limit, the $T$-matrix then takes the simple form

\begin{align}
    \mathcal{T}(\omega_n,\,\omega_n;{\bf r}_i)=\mathcal{V}(\omega_n,\,\omega_n;\,{\bf r}_i)+\mathcal{V}(\omega_n,\,\omega_n;{\bf r}_i)\mathcal{G}_{\textrm{hom}}(\omega_n;\,{\bf r}_i)\mathcal{T}(\omega_n,\,\omega_n;\,{\bf r}_i).
\end{align}
The $T$-matrix can be exactly solved for to yield
\begin{align}
    \mathcal{T}(\omega_n,\,\omega_n;\,{\bf r}_i)&=\bigg(\tau_0-\mathcal{V}(\omega_n,\,\omega_n;\,{\bf r}_i)\mathcal{G}_{\textrm{hom}}(\omega_n;\,{\bf r}_i)\bigg)^{-1}\mathcal{V}(\omega_n,\,\omega_n;\,{\bf r}_i)\notag\\
    &=\bigg(\mathcal{V}^{-1}(\omega_n,\,\omega_n;\,{\bf r}_i)-\mathcal{G}_{\textrm{hom}}(\omega_n;\,{\bf r}_i)\bigg)^{-1}.
\end{align}

\noindent In the case of time-translation invariance, the above potential becomes independent of frequency, and thus approaches a static limit $\mathcal{V}(\omega_n,\,\omega_n;\,{\bf r}_i)\equiv \mathcal{V}({\bf r}_i)$. The $T$-matrix also becomes frequency independent; namely, $\mathcal{T}(\omega_n,\,\omega_n;\,{\bf r}_i)\equiv \mathcal{T}({\bf r}_i)$. The above can be simplified if we let $\mathcal{G}_{\textrm{hom}}(\omega_n;{\bf r}_i)=G_0\tau_0+F_1\tau_1$ and $\mathcal{V}(\omega_n,\,\omega_n;\,{\bf r}_i)=V_0\tau_0+V_1\tau_1 +V_3\tau_3$. 
% Therefore, the above is simplified to
% \begin{align}
%     \mathcal{T}({\bf r}_i)&=\bigg[(V^{-1}_0-G_0)\tau_0+(V_1^{-1}-F_1)\tau_1+V_3^{-1}\tau_3\bigg]^{-1}\notag\\
%     &=\dfrac{(V_0^{-1}-G_0)\tau_0-(V_1^{-1}-F_1)\tau_1-V_3^{-1}\tau_3}{(V_0^{-1}-G_0)^2-(V_1^{-1}-F_1)^2-V_3^{-2}}.
% \end{align}
The form of the $T$-matrix for a $\tau_0$ impurity is then found by setting $\mathcal{V}({\bf r}_i)=V_0\tau_0$, while the $T$-matrix for the $\tau_3$ impurity can be found by setting $\mathcal{V}({\bf r}_i)=V_3\tau_3$.

We will focus now on the pole structure for a $\tau_0$ impurity. For the time being, we will remain agnostic to the form of the potential, and take $\mathcal{V}(\omega_n,\,\omega_n;\,{\bf r})=V_0\tau_0$. The poles then result in solutions where
\begin{align}
    (V_0^{-1}-G_0)^2-F_1^2=0.
\end{align}
Putting in the full form of the homogeneous Green's functions, we have a closed form for the frequencies $\omega^*$ where the pole in the $T$-matrix occurs:
\begin{align}
    \bigg(V_0^{-1}+\pi N(0)\dfrac{\omega^*}{\sqrt{\Delta^2-{\omega^*}^2}}\bigg)^2-\bigg(\pi N(0)\dfrac{\Delta}{\sqrt{\Delta^2-{\omega^*}^2}}\bigg)^2=0.
\end{align}
Solving for $\omega^*$, we find 
\begin{align}
    \omega^*=\pm \Delta_{\textrm{hom}}\bigg\{\dfrac{1-(\pi V_0N(0))^2}{1+(\pi V_0N(0))^2}\bigg\}.
\end{align}
Up until now, we have not specified the form of $V_0$, which is the lowest-order potential term in the $T$-matrix expansion for a $\tau_0$ impurity. In the next section, we will specialize to the case of the effective potential from a single electron scattering off a dynamical impurity.

\subsection{Local potential for the dynamical impurity}

As mentioned in the main text, the conclusions of this work is mostly agnostic to the form of $V_0$. This term may be any local term which causes a local particle-hole symmetric shift of the energy, and can be argued via symmetry constraints~\cite{Slager2015Aug}. Likewise, it also may be argued that a similar term may arise through a formal double-well potential via the isospin representation if we ignore the level splitting and transition amplitude~\cite{He2025Jun}. However, for our case, we can make material-specific estimates if we can make a connection to microscopics independent of specific internal TLS parameters, which we motivate via using the electron-boson interaction.

In general, we will assume a local interaction between a single electron and a localized dynamical defect. As in the scenario of isotropic impurities or vibrational impurities in the Anderson model~\cite{Hewson2001Dec,Al-Eryani2026Apr}, the scattering potential is taken to be of the form $g^2 D(i\omega_n - i\omega_{n'})$, where $D(i\omega_n-i\omega_m)$ is the free boson propagator evaluated at the energy transferred to the defect, i.e., the difference of the incoming and outgoing
electron Matsubara frequencies. A single electron scattering off this defect emits and reabsorbs (or absorbs and
re-emits) a single boson, which is instantaneous at the lowest order of the $T$-matrix. As in the case of two electrons scattering off the induced bosonic condensate, the effective potential includes the interaction vertices~\cite{Lipavsky2008Dec,Sopik2011Sep}, and thus should scale as a local (potential) energy shift ($\tau_0$). As such, this term is mathematically identical to the potential felt between electrons in Eliashberg theory after integrating out the phonons, where boson exchange between electrons across the Fermi sea mediates
Cooper pairing. Here, the same effective interaction instead describes a single electron scattering elastically
(or inelastically) off one localized dynamical defect, as in the case of localized bosons via isotopic disorder in 1D chains~\cite{Denisov2024Jul}.

The approximation of the interaction potential we take is similar to that of considered in the case of polarons, by which a single electron-boson interaction produces a constant energy shift proportional to $g^2$ and inversely proportional to the boson frequency~\cite{Dai2025Dec}. As a microscopic example, we can consider the $\tau_0$ as emerging from a polaron-like potential emerging from a local boson (the vibron) coupled to a charge density term. In this way, the fermionic Hamiltonian is modified by a bosonic term which goes as $H_b$, with a spin-independent Holstein coupling $g$:
\begin{align}
    H_{b}=\omega_0b_r^\dagger b_r+g(b_r+b_r^\dagger)\Psi_r^\dagger \tau_3 \Psi_r.
\end{align}
\noindent In the above, we have assumed a single boson at site $r$, and the Nambu spinor is now given in real space by $\Psi=\begin{pmatrix} c_\uparrow^\dagger & c_\downarrow\end{pmatrix}^T$. Therefore, to the lowest order of the $T$-matrix, the electron absorbs and emits a phonon simultaneously while on the local impurity site. Taking a Lang-Frisov transformation~\cite{Hohenadler2007}, we then remove the boson coupling and obtain an effective potential term:

\begin{align}
    H_b=\omega_0\left(b^\dagger +\dfrac{g}{\omega_0}\Psi_r^\dagger \tau_3 \Psi_r\right)\left(b +\dfrac{g}{\omega_0}\Psi_r^\dagger \tau_3 \Psi_r\right)-\dfrac{g^2}{\omega_0}\left(\Psi_r^\dagger \tau_3 \Psi_r\right)^2.
\end{align}

Defining new shifted bosonic operators, we now focus on the new term which is quadratic in $\Psi_r^\dagger \tau_3 \Psi_r$. Note that

\begin{align}
    \Psi_r^\dagger \tau_3 \Psi_r=\begin{pmatrix}c_\uparrow^\dagger & c_\downarrow \end{pmatrix}\begin{pmatrix} 1 & 0 \\ 0 & -1\end{pmatrix} \begin{pmatrix}c_\uparrow \\ c_\downarrow^\dagger \end{pmatrix}=n_\uparrow +n_\downarrow -1.
\end{align}

\noindent Therefore,

\begin{align}
    (\Psi^\dagger \tau_3 \Psi)^2&=(n_\uparrow +n_\downarrow -1)^2\notag\\
    &=(n_\uparrow +n_\downarrow)^2-2(n_\uparrow +n_\downarrow)+1\notag\\
    &\equiv n_{\textrm{tot}}(n_{\textrm{tot}}-2)+1\notag\\
    &\equiv (\Psi_r^\dagger \tau_3\Psi_r+1)(\Psi_r^\dagger \tau_3\Psi_r-1)+\Psi_r^\dagger\tau_0\Psi.
\end{align}
If $n_{\textrm{tot}}=0$ or $n_{\textrm{tot}}=2$, the above reduces to a pure $\tau_0$ term, while if $n_{\textrm{tot}}=1$, the above results in zero. As a result, in the elastic scattering limit, we have shown that the dynamical impurity at the lowest diagrammatic level appears as a $\tau_0$ energy shift. Physically, note that $n_\uparrow=n_\downarrow=1$ at the impurity site might be realized if the electron-phonon coupling is strong enough to induce a bipolaron state~\cite{Marsiglio1991Mar,Combescot1995May,Chubukov2026Apr}. Consequently, while the $\tau_0$ potential may be argued to emerge from some general energy inhomogeneity, microscopically such a potential may result from structural defects that induce localized low-frequency modes, which will in-turn be correlated with enhanced electron-phonon coupling~\cite{Tyner2025Oct,Heath2026Jan,Heath2026May}.   

Following from the above, we will assume an effective potential of $V_0\sim g^2 D(i\omega_n-i\omega_m)$ for the electron scattering off a dynamical impurity, with the propagator given by

\begin{align}
    D(i\omega_n-i\omega_\ell;\,{\bf r}_i)&
    =\int_0^\infty d\nu\,B(\nu)\dfrac{2\nu}{(\omega_n-\omega_\ell)^2+\nu^2}
    \notag\\
    &=\int_0^\infty d\nu \dfrac{\alpha^2 F(\nu)}{N(0)g^2}\dfrac{2\nu}{(\omega_n-\omega_\ell)^2+\nu^2}\tau_0\notag\\
    &=\dfrac{2}{N(0)g^2}\int_0^\infty d\nu\,\alpha^2 F(\nu)\dfrac{\nu}{(\omega_n-\omega_\ell)^2+\nu^2}\notag\\
    &=\dfrac{\lambda(i\omega_n-i\omega_m;\,{\bf r}_i)}{N(0)g^2},
    \end{align}
    where we have defined
    \begin{align}
        \lambda(i\omega_n-i\omega_m;\,{\bf r}_i)=\int_0^\infty d\nu \,\alpha^2 F(\nu;\,{\bf r}_i)\dfrac{2\nu}{(\omega_n-\omega_m)^2+\nu^2}.
    \end{align}
 \begin{figure}[t]
\includegraphics[width=0.85\columnwidth]{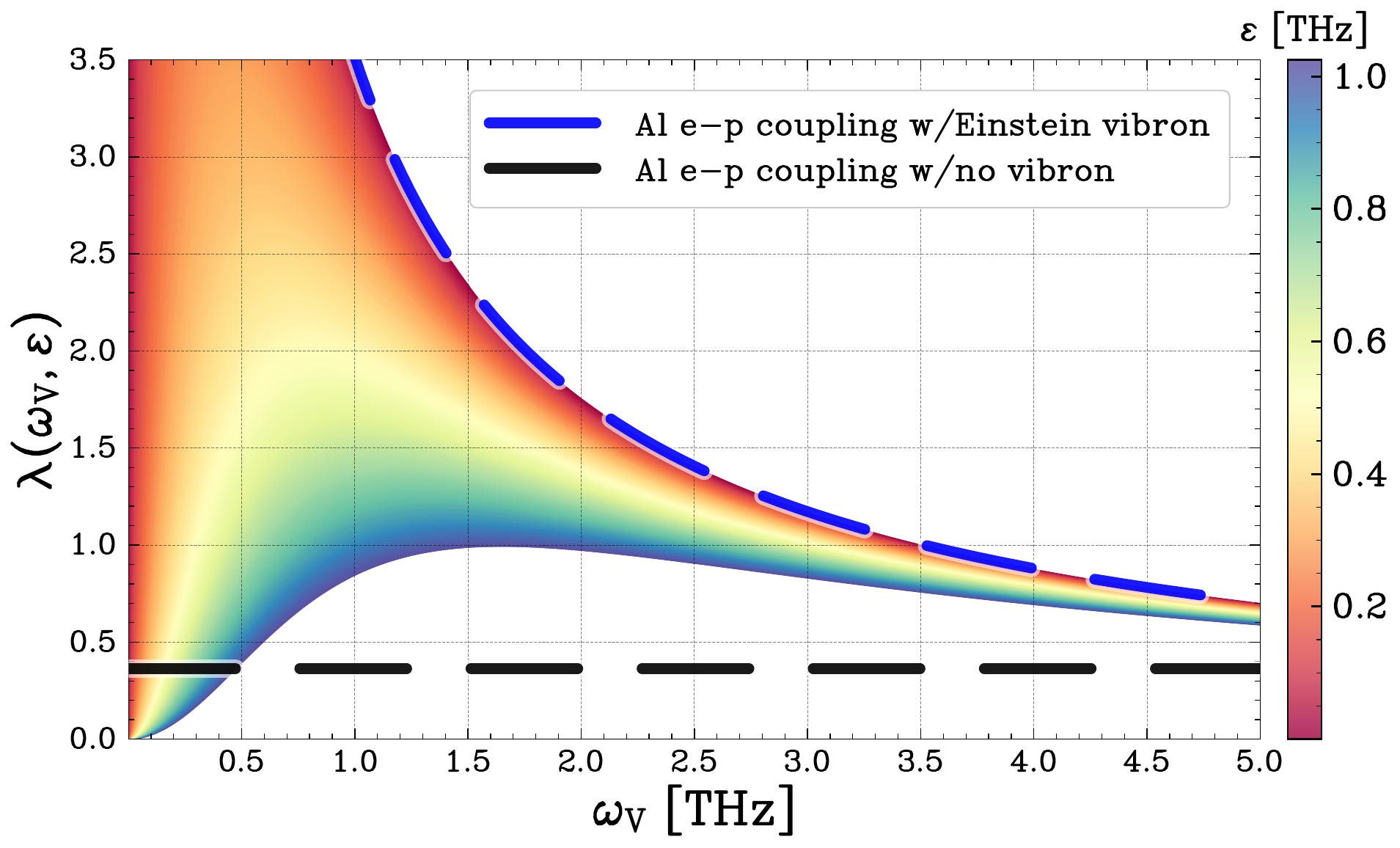}
\caption{\small A plot of the electron-boson coupling strength for a local vibron, using the aluminium value $\lambda=0.43$ as the large-frequency limit. For vanishing vibron frequency, the electron-boson coupling goes to zero, and thus we approach the static limit. In the limit of $\epsilon\rightarrow 0$, we obtain the Einstein boson result.
\label{fig:lambda}}
\end{figure}
    
As a consequence, we can say that
\begin{align}
V_0(i\omega_n,\,i\omega_m)=\dfrac{\lambda(i\omega_n-i\omega_m;\,{\bf r}_i)}{N(0)}.
\end{align}
For the case of a static impurity, the above reduces to $V_0(i\omega_n,\,i\omega_n;\,{\bf r}_i)=\lambda(0;\,{\bf r}_i)/N(0)$. Note that in the case of taking an Einstein boson as the dynamical impurity, the electron-boson coupling becomes inversely proportional to the boson frequency, and thus the potential apparently diverges in the static limit. This may be remedied by including a term $\epsilon$ in the spectral function $\alpha^2 F(\nu)$, thereby producing
Lorentzian approximation~\cite{Marsiglio2020Jun} to the vibron spectral function:
\begin{align}
    \alpha^2 F_V(\nu)=g^2 N(0)\dfrac{1}{\pi}\bigg\{\dfrac{\epsilon}{(\nu-\omega_V)^2+\epsilon^2}-\dfrac{\epsilon}{\omega_V^2+\epsilon^2}\bigg\}\Theta(\omega_V-|\nu-\omega_V|).
\end{align}
    
As a consequence, the electron-boson coupling becomes

\begin{align}
    \lambda_V(i\omega_n-i\omega_m;\,{\bf r}_i)=\dfrac{2g^2 N(0)}{\pi}\int_0^{2\omega_V}d\nu \,\bigg\{\frac{\nu}{(\omega_n-\omega_m)^2+\nu^2}\bigg\}\cdot\left\{\dfrac{\epsilon}{(\nu-\omega_V)^2+\epsilon^2}-\dfrac{\epsilon}{\omega_V^2+\epsilon^2}\right\}.
\end{align}
From which the form of $V(i\omega_n,\,i\omega_m)$ given in the main text immediately follows. In the static limit, note that we obtain

\begin{align}
    V_0(i\omega_n,\,i\omega_n;\,{\bf r}_i)= \dfrac{2g^2}{\pi}\int_0^{2\omega_V} \dfrac{d\nu}{\nu}\bigg\{\dfrac{\epsilon}{(\nu-\omega_V)^2+\epsilon^2}-\dfrac{\epsilon}{\omega_V^2+\epsilon^2}\bigg\},
\end{align}
and thus we may define in the static limit
\begin{align}
    \lambda_V(\omega_V,\,\epsilon)\equiv \lambda_V(i\omega_n,\,i\omega_n;\,{\bf r}_i)=\dfrac{2g^2 N(0)}{\pi}\int_0^{2\omega_V}\dfrac{d\nu}{\nu}\cdot\left\{\dfrac{\epsilon}{(\nu-\omega_V)^2+\epsilon^2}-\dfrac{\epsilon}{\omega_V^2+\epsilon^2}\right\}.
\end{align}

\noindent From the above, we note that in the limit of $\omega_V\rightarrow 0$ for any non-zero $\epsilon$, $V(i\omega_n,\,i\omega_n;\,{\bf r}_i)\rightarrow 0$, thus we recover the correct static limit (i.e., a vanishing of the $\tau_0$-term in the general impurity potential). 

From the form of the elastic-scattering limit $\lambda_V(i\omega_n,\,i\omega_n;\,{\bf r}_i)$, recall that we might obtain an estimate for the vibron's electron-boson coupling (and, thus, the effective potential for a dynamical impurity) for a real material if we rewrite the full electron-boson coupling in terms of the Einstein phonon result $\lambda_E=2 N(0)g^2/\omega_E$. We can then approximate $\lambda_E$ as the electron-phonon coupling strength of the material, and use the log average phonon frequency as an approximate $\omega_E$. In the limit of $\omega_V\gg 1$ THz, the local electron-boson coupling then approaches the material's baseline electron-phonon coupling.  In Fig. 1, we see the local electron-boson coupling for the base material of aluminium (Al). As $\omega_V\rightarrow 0$, we approach the static limit on the impurity site, and thus the electron-boson coupling goes to zero. As $\epsilon\rightarrow 0$, we obtain the Einstein phonon result. A similar approximation is used to create Table II in the main text, with $\epsilon$ minimized so that the maximum of the electron-boson coupling reaches the maximum possible value allowed by Eliashberg theory over all $\omega_V$.

\subsection{Subgap pole structure in the inelastic scattering limit}

The previous discussion focused on the form of the subgap pole in the case of elastic scattering. However, we want to check our results for the case of inelastic scattering. In the elastic limit, the impurity potential is diagonal in Matsubara frequency, and thus $\mathcal{V}(i\omega_n,\,i\omega_m;\,{\bf r}_i)=\mathcal{V}({\bf r}_i)\delta_{nm}$, and thus the $T$-matrix equation reduces to a closed form. This exact closed form solution fails once we consider the internal dynamics of the TLS or vibron, in which case there is a possibility that the incoming frequency $\omega_n$ might not be equivalent to the outgoing frequency $\omega_m$. In such a case,  $\mathcal{V}(i\omega_n,\,i\omega_m;\,{\bf r}_i)$ becomes off-diagonal in the Matsubara frequencies. If we are to include the off-diagonal contributions, the $T$-matrix equation now becomes an equation over all Matsubara frequencies, and in evaluating $T(i\omega_n,\,i\omega_m)$ for the physical frequency pair, we sum over all intermediate legs $\omega_\ell$~\cite{Bruus2004Sep}:

\begin{align}
    \mathcal{T}(\omega_n,\,\omega_m;{\bf r}_i)=\mathcal{V}(\omega_n,\,\omega_m;{\bf r}_i)+\sum_{\ell}\mathcal{V}(\omega_n,\,\omega_\ell;{\bf r}_i)\mathcal{G}_{\textrm{hom}}(\omega_\ell;\,{\bf r}_i)\mathcal{T}(\omega_\ell,\,\omega_m;{\bf r}_i).
\end{align}

\noindent These off-diagonal terms are the off-shell scattering events, which manifest as intermediate energy-exchange events in the Matsubara sum. We can recast the above equation as a new matrix equation, with the full $T$-matrix split into a $\tau_0$ (normal channel) and $\tau_1$ (anomalous channel) contributions:

\begin{align}
\mathcal{T}(\omega_n,\,\omega_m;\,{\bf r}_i)&=\mathcal{V}(\omega_n,\,\omega_m;\,{\bf r}_i)+\sum_\ell \mathcal{V}(\omega_n,\,\omega_\ell;\,{\bf r}_i) \mathcal{G}_\textrm{hom}(\omega_\ell,\,\omega_\ell;\,{\bf r}_i) \mathcal{T}(\omega_\ell,\,\omega_m;\,{\bf r}_i)\notag\\
\rightarrow 
    \mathcal{T}_{nm}
    &=\mathcal{V}_{nm}+\sum_\ell \mathcal{V}_{n\ell} \mathcal{G}_{\ell\ell}\mathcal{T}_{\ell m}\notag\\
    &=V_{nm}\tau_0+\dfrac{1}{\beta}\sum_\ell V_{n\ell}\tau_0  (G_{\ell\ell}\tau_0+F_{\ell\ell}\tau_1)(T_{\ell m,\,+}\tau_0+T_{\ell m,\,-}\tau_1)\notag\\
        % &=V_{nm}\tau_0+\dfrac{1}{\beta}\sum_\ell V_{n\ell}  (G_\ell\tau_0-F_\ell\tau_1)(T_{\ell m}^0\tau_0+T_{\ell m}^1\tau_1)\notag\\
        &=V_{nm}\tau_0+\sum_\ell V_{n\ell}  (G_{\ell\ell} T_{\ell m,\,+}\tau_0+G_{\ell\ell} T_{\ell m,\,-} \tau_1 +F_{\ell\ell} T_{\ell m,\,+} \tau_1 +F_{\ell\ell} T_{\ell m,\,-} \tau_0)\notag\\
        &=\tau_0\bigg\{V_{nm}+\sum_\ell \bigg(V_{n\ell}G_{\ell\ell} T_{\ell m,\,+} +V_{n\ell} F_{\ell\ell} T_{\ell m,\,-} \bigg)\bigg\}
        +\tau_1 \bigg\{\sum_\ell \bigg(V_{n\ell} G_{\ell\ell} T_{\ell m,\,-}+V_{n\ell} F_{\ell\ell} T_{\ell m,\,+}\bigg) \bigg\}.
\end{align}

\noindent As such, we have written $\mathcal{T}(\omega_n,\,\omega_m;\,{\bf r}_i)\equiv \mathcal{T}_{nm}$ as elements of the matrix ${\bf T}$, with

\begin{align}
    {\bf T}&={\bf T}_+\otimes \tau_0+{\bf T}_-\otimes \tau_1.
    %\notag\\
    %T_{nm}&=T_{nm}^{(0)}\tau_0+T_{nm}^{(1)}\tau_1
\end{align}

\noindent Note that we take the new notation where a matrix in bold is now written as a matrix over the Matsubara indices, as described in the text. Similarly, we see that in the above we have re-written the potential and the homogeneous Green's functions as matrices over the Matsubara frequencies as well. However, as the Green's functions are homogeneous, they are purely diagonal. Also note that the index $i$ of the matrix and the Matsubara index is given by $\ell$ are related by $j=\ell+N$, so that $\ell=-N$ corresponds to $j=0$. 

The equations for the components of the T-matrix above are given by
\begin{align}
    T_{nm,\,+}=V_{nm}+\sum_\ell \bigg(V_{n\ell}G_{\ell\ell} T_{\ell m,\,+}+V_{n\ell}F_{\ell\ell} T_{\ell m,\,-}\bigg),
\end{align}
\begin{align}
T_{nm,\,-}=\sum_\ell \bigg(V_{n\ell} G_{\ell\ell} T_{\ell m,\,-}+V_{n\ell} F_{\ell\ell} T_{\ell m,\,+}\bigg).
\end{align}

\noindent These can be rewritten in the form

\begin{align}
    {\bf T}_+={\bf V}\bigg\{{\bf 1}+\left( {\bf G}_{\textrm{hom}}{\bf T}_++{\bf F}_{\textrm{hom}}{\bf T}_-\right)\bigg\},
\end{align}

\begin{align}
{\bf T}_-={\bf V}\left({\bf G}_{\textrm{hom}}{\bf T}_-+{\bf F}_{\textrm{hom}}{\bf T}_+\right).
\end{align}

From the above, we can now solve for ${\bf T}_+$ and ${\bf T}_-$. From the two equations above, we can rewrite these as 

\begin{align}
    \bigg({\bf 1}-{\bf V}{\bf G}_{\textrm{hom}}\bigg){\bf T}_+-{\bf V}{\bf F}_{\textrm{hom}}{\bf T}_-={\bf V},
\end{align}
\begin{align}
    \bigg({\bf 1}-{\bf V}{\bf G}_{\textrm{hom}}\bigg){\bf T}_--{\bf V}{\bf F}_{\textrm{hom}}{\bf T}_+=0.
\end{align}

Adding and subtracting the above, we obtain

\begin{align}
    \bigg({\bf 1}-\dfrac{1}{\beta}{\bf V}{\bf G}_{\textrm{hom}}-\dfrac{1}{\beta}{\bf V}{\bf F}_{\textrm{hom}}\bigg){\bf T}_++\bigg(1-\dfrac{1}{\beta}{\bf V}{\bf G}_{\textrm{hom}}-\dfrac{1}{\beta}{\bf V}{\bf F}_{\textrm{hom}}\bigg){\bf T}_-={\bf V},
\end{align}
\begin{align}
    \bigg({\bf 1}-\dfrac{1}{\beta}{\bf V}{\bf G}_{\textrm{hom}}+\dfrac{1}{\beta}{\bf V}{\bf F}_{\textrm{hom}}\bigg){\bf T}_+-\bigg(1-\dfrac{1}{\beta}{\bf V}{\bf G}_{\textrm{hom}}+\dfrac{1}{\beta}{\bf V}{\bf F}_{\textrm{hom}}\bigg){\bf T}_-={\bf V}.
\end{align}

\noindent These can be solved to yield 

\begin{align}
    {\bf T}_++{\bf T}_-=\bigg({\bf 1}-\dfrac{1}{\beta}{\bf V}{\bf G}_{\textrm{hom}}-\dfrac{1}{\beta}{\bf V}{\bf F}_{\textrm{hom}}\bigg)^{-1}{\bf V},
\end{align}
\begin{align}
    {\bf T}_+-{\bf T}_-=\bigg({\bf 1}-{\bf V}{\bf G}_{\textrm{hom}}+{\bf V}{\bf F}_{\textrm{hom}}\bigg)^{-1}{\bf V}.
\end{align}

\noindent Now, we will rewrite ${\bf T}_+$ and ${\bf T}_-$ to solve in each term individually:

\begin{align}
    {\bf T}_+&=\dfrac{1}{2}\bigg\{\bigg({\bf T}_++{\bf T}_-\bigg)+\bigg({\bf T}_+-{\bf T}_-\bigg)\bigg\}\notag\\
    &=\dfrac{1}{2}\bigg\{\bigg({\bf 1}-{\bf V}{\bf G}_{\textrm{hom}}-{\bf V}{\bf F}_{\textrm{hom}}\bigg)^{-1}+\bigg({\bf 1}-{\bf V}{\bf G}_{\textrm{hom}}+{\bf V}{\bf F}_{\textrm{hom}}\bigg)^{-1}\bigg\}{\bf V},
    % \notag\\
    % &=\dfrac{1}{2}{\bf V}^{-1}\bigg\{\bigg({\bf V}^{-1}-\dfrac{1}{\beta}\bigg\{{\bf G}+{\bf F}\bigg\}\bigg)^{-1}+\bigg({\bf V}^{-1}-\dfrac{1}{\beta}\bigg\{{\bf G}-{\bf F}\bigg\}\bigg)^{-1}\bigg\}{\bf V}
\end{align}
\begin{align}
    {\bf T}_- &=\dfrac{1}{2}\bigg\{\bigg({\bf T}_++{\bf T}_-\bigg)-\bigg({\bf T}_+-{\bf T}_-\bigg)\bigg\}\notag\\
    &=\dfrac{1}{2}\bigg\{\bigg({\bf 1}-{\bf V}{\bf G}_{\textrm{hom}}-{\bf V}{\bf F}_{\textrm{hom}}\bigg)^{-1}-\bigg({\bf 1}-{\bf V}{\bf G}_{\textrm{hom}}+{\bf V}{\bf F}_{\textrm{hom}}\bigg)^{-1}\bigg\}{\bf V}.
    % \notag\\
    % &=\dfrac{1}{2}{\bf V}^{-1}\bigg\{\bigg({\bf V}^{-1}-\dfrac{1}{\beta}\bigg\{{\bf G}+{\bf F}\bigg\}\bigg)^{-1}-\bigg({\bf V}^{-1}-\dfrac{1}{\beta}\bigg\{{\bf G}-{\bf F}\bigg\}\bigg)^{-1}\bigg\}{\bf V}
\end{align}

\noindent We will rewrite this in the form
\begin{align}
    {\bf T}_\pm=\dfrac{1}{2}\bigg\{{\bf A}_-^{-1}\pm{\bf A}_+^{-1}\bigg\}{\bf V},
\end{align}
where
\begin{align}
    {\bf A}_{\pm}={\bf 1}-{\bf V}\bigg\{{\bf G}_{\textrm{hom}}\pm {\bf F}_{\textrm{hom}}\bigg\}.
\end{align}

\noindent This reproduces the form of the equations given in the main text. 

 \begin{figure}[t]
\includegraphics[width=0.85\columnwidth]{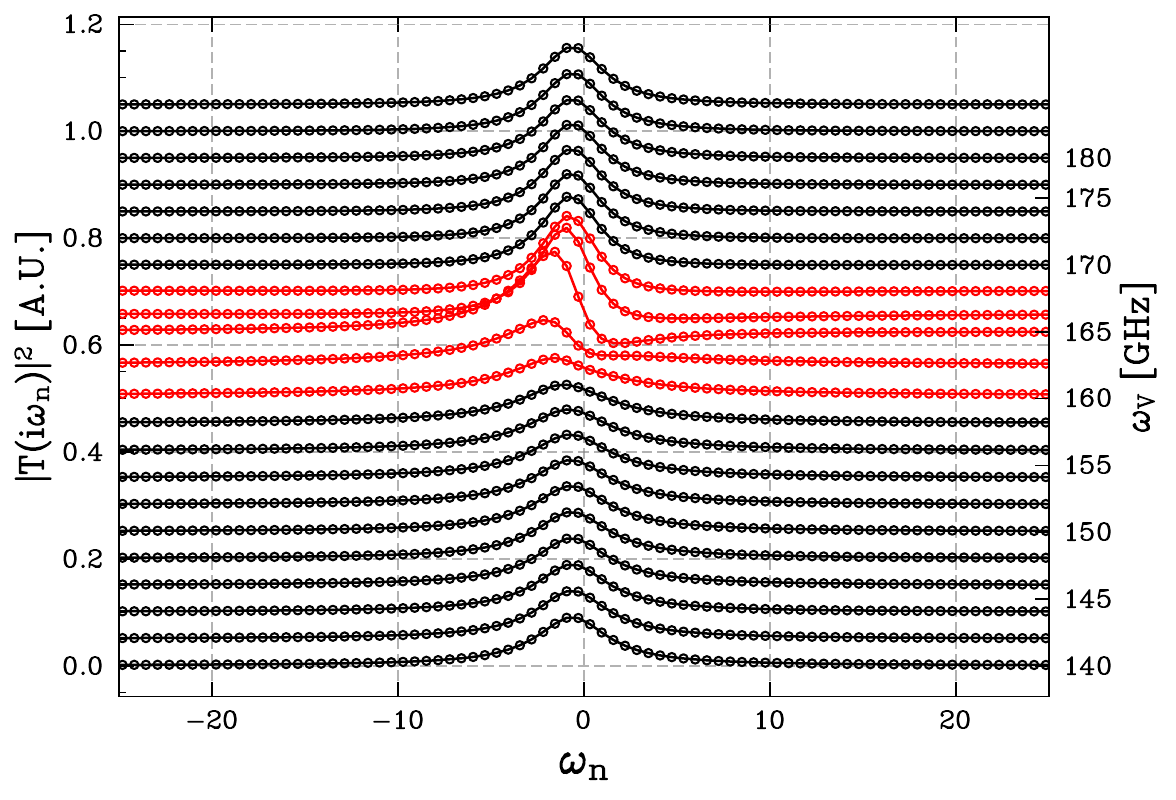}
\caption{\small The value of $|T_-(i\omega_n)|^2$ vs $\omega_n$ for several values of the vibron frequency $\omega_V$. The same parameters for the aluminium plot are used here, with $\epsilon=500$ GHz taken. Note that we take the diagonal component of ${\bf T}_-$, as nearly every other value off the diagonal is zero. For vibron frequencies $\omega_V\sim 160-170$ GHz, the Lorentzian structure breaks down. This vibron frequency regime corresponds roughly to the same regime where the pole approaches zero on the real frequency axis. 
\label{fig:Appendix2}}
\end{figure}

We have shown in the main text that the potential ${\bf V}$ becomes diagonal dominant for values of the vibron frequency $\omega_V$ below around $100$ GHz, thus suggesting that the elastic scattering approximation remains accurate for vibrons in the low sub-THz regime. To further understand the pole structure in the inelastic limit, we can infer qualitative behavior by considering how $|{\bf T}_-|^2$ changes upon shifting $\omega_V$. As the $T$ matrix is mostly dominantly diagonal on the Matsubara frequency axis, we can consider the behavior of $T_-^2(i\omega_n)$. The pole emerging in the real-axis equations should appear as a Lorentzian when plotting ${\bf T}_-^2$. In Fig.~\ref{fig:Appendix2}, we plot values of $|T_-(i\omega_n)|^2$ vs the Matsubara frequency for several vibron frequencies, taking material-specific values of Al where appropriate (i.e., in the estimate of the vibron potential) and $\epsilon=500$ GHz. Around $\omega_V\sim 160-170$ GHz, we see a breakdown of the Lorentzian structure (shown in red). On the real axis for the same vibron parameters, note that the subgap poles reaches zero around $\omega_V\approx 150$ GHz. As such, the breakdown of the Lorentzian structure seen in Fig.~\ref{fig:Appendix2} can be interpreted as the $T$-matrix pole momentarily disappearing at zero frequency before increasing again. While this approach for solving for the $T$-matrix pole does not yield a close quantitative form for the subgap bound state, it confirms our elastic scattering approximation used throughout the main text.

\section{Local modulation of the density of states: the $T$-matrix for $|\omega|>\Delta$}

\subsection{The homogeneous and local superconducting density of states}
The total density of states of the superconductor in the homogeneous limit is defined as
\begin{align}
    N_{\textrm{hom}}(\omega)=-\dfrac{1}{\pi}\sum_k \Im\bigg[\mathcal{G}_{\textrm{hom}}(\omega,\,k)\bigg]_{11}.
\end{align}
In a BCS superconductor, we might write the homogeneous Green's function in the Nambu representation as
\begin{align}
\mathcal{G}_{\textrm{hom}}(\omega,k)&=\dfrac{\omega \tau_0+\xi_k \tau_3+\Delta_k \tau_1}{\omega^2-\xi_k^2-\Delta_k^2},
%\notag\\
  %  &=\dfrac{\omega\tau_0+\xi_k\tau_3}{\omega^2-\xi_k^2-\Delta_k^2}+\dfrac{\Delta_k \tau_1}{\omega^2-\xi_k^2-\Delta_k^2}
\end{align}
\noindent  Taking the upper term on the diagonal, we then only retain the normal component:

\begin{align}
    \mathcal{G}_{11}(\omega,k)&=\dfrac{\omega+\xi_k}{(\omega+i\delta)^2-\xi_k^2-\Delta_k^2}\notag\\
    % &\equiv \dfrac{\omega+\xi_k}{(\omega+i\delta_k)^2-E_k^2},\qquad E_k\equiv \sqrt{\xi_k^2+\Delta_k^2}
    % \notag\\
    % &=\dfrac{1}{2}\bigg(1+\dfrac{\xi_k}{E_k}\bigg)\dfrac{1}{\omega+i\delta-E_k}+\dfrac{1}{2}\bigg(1-\dfrac{\xi_k}{E_k}\bigg)\dfrac{1}{\omega+i\delta_k+E_k}\notag\\
    % &\equiv \dfrac{u_k^2}{\omega+i\delta-E_k}+\dfrac{v_k^2}{\omega+i\delta_k+E_k}\notag\\
    &= \bigg\{P\dfrac{u_k^2}{\omega-E_k}-u_k^2i\pi \delta(\omega-E_k) \bigg\}+\bigg\{P\dfrac{v_k^2}{\omega+E_k}-v_k^2i\pi \delta(\omega+E_k)\bigg\}.
\end{align}
\noindent where in the last line we have invoked the Sokhotski–Plemelj theorem and used
\begin{align}
    E_k\equiv \sqrt{\xi_k^2+\Delta_k^2},\qquad u_k^2\equiv\dfrac{1}{2}\bigg(1+\dfrac{\xi_k}{E_k}\bigg),\qquad v_k^2\equiv\dfrac{1}{2}\bigg(1-\dfrac{\xi_k}{E_k}\bigg).
\end{align}

\noindent Focusing on $\omega>0$ for the time being, we have 
\begin{align}
\mathcal{G}_{11}(\omega,\,k)=P\dfrac{u_k^2}{\omega-E_k}-u_k^2i\pi \delta(\omega-E_k) .
\end{align}
\noindent From the above, the imaginary part is given by

\begin{align}
    \Im \mathcal{G}_{11}(\omega,\,{\bf k})&=\Im \bigg[P\dfrac{u_k^2}{\omega-E_k}\bigg]-\Re [u_k^2]\pi \delta(\omega-E_k).
\end{align}

\noindent As we are dealing with BCS theory, $\Delta_k$ and $\xi_k$ should both be real. Therefore, we should be able to write 

\begin{align}
% \Im \mathcal{G}_{11}(\omega,{\bf k})&=- u_k^2 \pi \delta(\omega-E_k)\notag\\
% &
% =-\dfrac{\pi}{2} \bigg(1+\dfrac{\xi_k}{E_k}\bigg)\delta(\omega-E_k)\notag\\
%     \rightarrow N(\omega)&=\dfrac{1}{2}\sum_k\bigg(1+\dfrac{\xi_k}{E_k}\bigg)\delta(\omega-E_k)\notag\\
%     % &= \dfrac{1}{2}\int d\epsilon_k \, N_0 \bigg(1+\dfrac{\epsilon_k}{E_k}\bigg)\delta(\omega-E_k)\notag\\
%     &\rightarrow \dfrac{1}{2}\int d\xi_k \, N_0 \bigg(1+\dfrac{\xi_k}{E_k}\bigg)\delta(\omega-E_k)\notag\\
% \rightarrow 
\dfrac{N(\omega)}{N(0)}&=\dfrac{1}{2}\int d\xi_k \bigg(1+\dfrac{\xi_k}{E_k}\bigg)\delta(\omega-E_k).
\end{align}

\noindent Note that we can write the delta function as 
\begin{align}
    \delta(f(\epsilon_k))=\sum_i\dfrac{\delta(\xi_k-\xi_k^i)}{|f'(\xi^i)|},
\end{align}
where $\xi_k^i$ corresponds to zeros of the function $f(\xi_k)=\omega-\sqrt{\xi_k^2+\Delta_k^2}$. The poles of this function can easily be found to be $\xi_k^{\pm}=\pm \sqrt{\omega^2-\Delta_k^2}$, however note that we have the physical constraint that $\xi_k=\epsilon_k-\mu$ must be real. Therefore, we'll list the poles as $\xi_k^{\pm}=\pm \Re\sqrt{\omega^2-\Delta_k^2}$. The derivative of our function is
\begin{align}
    |f'(\xi_k)|=\bigg|\dfrac{d}{d\xi_k}f(\xi_k)\bigg|=\dfrac{|\xi_k|}{\sqrt{\xi_k^2+\Delta_k^2}}.
\end{align}
Thus, we find that 
\begin{align}
\delta(f(\xi_k))&=\sum_i \dfrac{\delta(\xi_k-\xi_k^i)}{|f'(\xi^i)|}=\dfrac{\omega}{|\xi_k^{\pm}|}\bigg\{\delta(\xi_k-\xi_k^+)
+\delta(\xi-\xi_k^-)\bigg\}.
\end{align}
\noindent Therefore, upon assuming a $k$-independent gap and now considering all $\omega$ without loss of generality, we find that the density of states becomes
\begin{align}
    \dfrac{N(\omega)}{N(0)}&=\dfrac{\omega}{\sqrt{\omega^2-\Delta^2}}\Theta(\omega-\Delta).
\end{align}

We will now consider the case of the local density of states. Assuming an impurity at site ${\bf r}={\bf r}_i$, the total density of states a distance ${\bf r}$ from the impurity is given by the following:

\begin{align}
N(\omega,\,{\bf r})&=N_{\textrm{hom}}(\omega)+\delta N(\omega,\,{\bf r})\notag\\
&=-\dfrac{1}{\pi}\Im \bigg[\mathcal{G}_{\textrm{hom}}(\omega,\,0)\bigg]_{11}-\dfrac{1}{\pi}\Im \bigg[\mathcal{G}_{\textrm{hom}}(\omega,\,{\bf r}-{\bf r}_i)\mathcal{T}(\omega,\,{\bf r}_i)\mathcal{G}_{\textrm{hom}}(\omega,\,{\bf r}_i-{\bf r})\bigg]_{11}\notag\\
&\equiv-\dfrac{1}{\pi}\Im \bigg[\mathcal{G}_{\textrm{hom}}(\omega)\bigg]_{11}-\dfrac{1}{\pi}\Im \bigg[\mathcal{G}_{\textrm{hom}}(\omega,\,{\bf R}_i)\mathcal{T}(\omega)\mathcal{G}_{\textrm{hom}}(\omega,\,-{\bf R}_i)\bigg]_{11}.
\end{align}
The second term is modified by the $T$-matrix, which can be solved for to ultimately find the LDOS modulation induced by some general impurity.

\subsection{LDOS in the presence of a static $\tau_3$-impurity}

We'll first consider the well-known $\tau_3$ impurity. Recall that

\begin{align}
    \mathcal{T}_3(\omega)&=\dfrac{-G_0\tau_0+F_1\tau_1-V_3^{-1}\tau_3}{G_0^2-F_1^2-V_3^{-2}}\notag\\
%    &=\dfrac{-G_0\tau_0+F_1\tau_1-V_3^{-1}\tau_3}{-(\pi N(0))^2-V_3^{-2}}\notag\\
    &=\dfrac{1}{-(\pi N(0))^2-V_3^{-2}}\bigg\{-G_0\tau_0+F_1\tau_1-V_3^{-1}\tau_3\bigg\}
\end{align}
With the above, we can readily solve for the defining term in the LDOS:

    \begin{align}
        &\bigg[\mathcal{G}_{\textrm{hom}}(\omega;\,{\bf R}_i)\mathcal{T}(\omega;\,0)\mathcal{G}_{\textrm{hom}}(\omega,\,-{\bf R}_i)\bigg]_{11}\notag\\
 %       &=\dfrac{\sin^2(k_F R_i)}{(k_F R_i)^2}e^{-R_i/\ell}\bigg[\mathcal{G}_{\textrm{hom}}(\omega;\,0)\mathcal{T}(\omega;\,0)\mathcal{G}_{\textrm{hom}}(\omega;\,0)\bigg]\notag\\
 %       &=\dfrac{\sin^2(k_F R_i)}{(k_F R_i)^2}e^{-R_i/\ell}\bigg[\bigg\{G_0\tau_0+F_1\tau_1\bigg\}\bigg\{\dfrac{-G_0\tau_0+F_1\tau_1-V_3^{-1}\tau_3}{-(\pi N(0))^2-V_3^{-2}}\bigg\}\bigg\{G_0\tau_0+F_1\tau_1\bigg\}\bigg]\notag\\
        &=\dfrac{\sin^2(k_F R_i)}{(k_F R_i)^2}e^{-R_i/\ell}\dfrac{1}{-(\pi N(0))^2-V_3^{-2}}
        \bigg[\bigg\{G_0\tau_0+F_1\tau_1\bigg\}\bigg\{-G_0\tau_0+F_1\tau_1-V_3^{-1}\tau_3\bigg\}\bigg\{G_0\tau_0+F_1\tau_1\bigg\}\bigg]_{11}
    \end{align}
\noindent For this expression, we can utilize the identity:
    \begin{align}
        [(a_1\tau_0+a_2\tau_1)(b_1\tau_0+b_2\tau_1+b_3\tau_3)(a_1\tau_0+a_2\tau_1)]_{11}=b_1(a_1^2+a_2^2)+2a_1a_2b_2+(a_1^2-a_2^2)b_3
    \end{align}
\noindent We therefore find that
\begin{align}
     &\bigg[\bigg\{G_0\tau_0+F_1\tau_1\bigg\}\bigg\{-G_0\tau_0+F_1\tau_1-V_3^{-1}\tau_3\bigg\}\bigg\{G_0\tau_0+F_1\tau_1\bigg\}\bigg]_{11}\notag\\
     &=-G_0(G_0^2+F_1^2)+2G_0F_1^2-V_3^{-1}(G_0^2-F_1^2)\notag\\
     %&=-G_0(G_0^2-F_1^2)-V_3^{-1}(G_0^2-F_1^2)\notag\\
     %&=-(G_0+V_3^{-1})(G_0^2-F_1^2)\notag\\
     &=(G_0+V_3^{-1})(\pi N(0))^2
\end{align}
and thus the main relation becomes
\begin{align}
    &\dfrac{1}{-(\pi N(0))^2-V_3^{-2}}
        \bigg[\bigg\{G_0\tau_0+F_1\tau_1\bigg\}\bigg\{-G_0\tau_0+F_1\tau_1-V_3^{-1}\tau_3\bigg\}\bigg\{G_0\tau_0+F_1\tau_1\bigg\}\bigg]_{11}\notag\\
        %&=-\dfrac{(\pi N(0))^2 (G_0+V_3^{-1})}{(\pi N(0))^2+V_3^{-2}}\notag\\
        &=-\dfrac{\pi^2 N^2(0)V_3^2}{\pi^2 N^2(0)V_3^2+1}\bigg(G_0+V_3^{-1}\bigg)
\end{align}
Recall the form of $G_0$:
\begin{align}
    G_0=-i\pi N(0)\dfrac{|\omega|}{\sqrt{\omega^2-\Delta^2}}
\end{align}
As the modulation in the LDOS is only proportional to the imaginary contribution, we obtain the following:
\begin{align}
   &-\dfrac{1}{\pi} \Im\bigg[\mathcal{G}_{\textrm{hom}}(\omega;\,{\bf R}_i)\mathcal{T}(\omega;\,0)\mathcal{G}_{\textrm{hom}}(\omega,\,-{\bf R}_i)\bigg]_{11}\notag\\
   %&=-\dfrac{1}{\pi}\dfrac{\sin^2(k_F R_i)}{(k_F R_i)^2}e^{-R_i/\ell}\Im \bigg[-\dfrac{\pi^2 N^2(0)V_3^2}{\pi^2 N^2(0)V_3^2+1}\bigg(-i\pi N(0)\dfrac{|\omega|}{\sqrt{\omega^2-\Delta^2}}+V_3^{-1}\bigg)\bigg]\notag\\
   %&=-N(0)\dfrac{|\omega|}{\sqrt{\omega^2-\Delta^2}}\dfrac{\sin^2(k_F R_i)}{(k_F R_i)^2}e^{-R_i/\ell}\dfrac{\pi^2 N^2(0)V_3^2}{\pi^2 N^2(0)V_3^2+1}\notag\\
   &=-N_{\textrm{hom}}(\omega)\dfrac{\sin^2(k_F R_i)}{(k_F R_i)^2}e^{-R_i/\ell}\dfrac{\pi^2 V_3^2 N^2(0)}{\pi^2 V_3^2 N^2(0)+1}
\end{align}
As such, for the $\tau_3$ impurity, we find that~\cite{Zarea2023Oct}
\begin{align}
    N(\omega;\,{\bf R}_i)=N_{\textrm{hom}}(\omega)\left\{1-\dfrac{\sin^2(k_F R_i)}{(k_F R_i)^2}e^{-R_i/\ell}\dfrac{\pi^2 V_3^2 N^2(0)}{\pi^2 V_3^2 N^2(0)+1}\right\}
\end{align}

\subsection{LDOS in the presence of a dynamical $\tau_0$-impurity}

We'll now turn our attention to the $\tau_0$ impurity. For the $\tau_0$-impurity, we have

\begin{align}
    \mathcal{T}_0(i\omega_n)=
    \dfrac{(V_0^{-1}-G_0)\tau_0+F_1\tau_1}{(V_0^{-1}-G_0)^2-F_1^2}
\end{align}

\noindent We proceed for this form of the impurity as we did for the $\tau_3$ variation:

    \begin{align}
        &\bigg[\mathcal{G}_{\textrm{hom}}(\omega;\,{\bf R}_i)\mathcal{T}(\omega;\,0)\mathcal{G}_{\textrm{hom}}(\omega,\,-{\bf R}_i)\bigg]_{11}\notag\\
      %  &=\dfrac{\sin^2(k_F R_i)}{(k_F R_i)^2}e^{-R_i/\ell}\bigg[\mathcal{G}_{\textrm{hom}}(\omega;\,0)\mathcal{T}(\omega;\,0)\mathcal{G}_{\textrm{hom}}(\omega;\,0)\bigg]\notag\\
      %  &=\dfrac{\sin^2(k_F R_i)}{(k_F R_i)^2}e^{-R_i/\ell}\bigg[\bigg\{G_0\tau_0+F_1\tau_1\bigg\}\bigg\{\dfrac{(V_0^{-1}-G_0)\tau_0+F_1\tau_1}{(V_0^{-1}-G_0)^2-F_1^2}\bigg\}\bigg\{G_0\tau_0+F_1\tau_1\bigg\}\bigg]\notag\\
        &=\dfrac{\sin^2(k_F R_i)}{(k_F R_i)^2}e^{-R_i/\ell}\dfrac{1}{(V_0^{-1}-G_0)^2-F_1^2}
        \bigg[\bigg\{G_0\tau_0+F_1\tau_1\bigg\}\bigg\{(V_0^{-1}-G_0)\tau_0+F_1\tau_1\bigg\}\bigg\{G_0\tau_0+F_1\tau_1\bigg\}\bigg]_{11}
    \end{align}

\noindent We'll now use the identity
\begin{align}
[(a_1\tau_0+a_2\tau_1)(b_1\tau_0+b_2\tau_1)(a_1\tau_0+a_2\tau_1)]_{11}=(a_1^2+a_2^2)b_1+2a_1a_2b_2
\end{align}

\noindent to write
\begin{align}
        &\bigg[\bigg\{G_0\tau_0+F_1\tau_1\bigg\}\bigg\{(V_0^{-1}-G_0)\tau_0+F_1\tau_1\bigg\}\bigg\{G_0\tau_0+F_1\tau_1\bigg\}\bigg]_{11}\notag\\
        % &=(G_0^2+F_1^2)(V_0^{-1}-G_0)+2G_0F_1^2\notag\\
        % &=(G_0^2+F_1^2)V_0^{-1}-(G_0^2+F_1^2)G_0+2G_0F_1^2\notag\\
        % &=(G_0^2+F_1^2)V_0^{-1}-G_0\left(G_0^2+F_1^2-2F_1^2\right)\notag\\
        % &=(G_0^2+F_1^2)V_0^{-1}-G_0\left(G_0^2-F_1^2\right)\notag\\
        &=(G_0^2+F_1^2)V_0^{-1}+G_0\left(F_1^2-G_0^2\right)
\end{align}

We now want to solve for the above and simplify. Note that the change in the local density of states is only proportional to the imaginary portion, and thus we will assume $|\omega|>\Delta$.
% \begin{align}
%     G_0=-i\pi N(0)\dfrac{|\omega|}{\sqrt{\omega^2-\Delta^2}},\qquad F_1 =-i\pi N(0)\dfrac{\Delta}{\sqrt{\omega^2-\Delta^2}}
% \end{align}
As such,
\begin{align}
    G_0^2+F_1^2=-\pi^2 N^2(0)\dfrac{\omega^2+\Delta^2}{\omega^2-\Delta^2}\equiv -(\pi N(0))^2\dfrac{\bar{\omega}^2+1}{\bar{\omega}^2-1}
\end{align}
\begin{align}
    F_1^2-G_0^2=(\pi N(0))^2
\end{align}
\noindent Thus, the term given above becomes
\begin{align}
    &\phantom{=}-(\pi N(0))^2 V_0^{-1}\dfrac{\bar{\omega}^2+1}{\bar{\omega}^2-1}+G_0(\pi N(0))^2=(\pi N(0))^2\bigg\{-V_0^{-1}\dfrac{\bar{\omega}^2+1}{\bar{\omega}^2-1}+G_0\bigg\}
\end{align}
\noindent Now, we'll simplify the denominator:
\begin{align}
    (V_0^{-1}-G_0)^2-F_1^2&=V_0^{-2}-2V_0^{-1}G_0+G_0^2-F_1^2=V_0^{-2}-2V_0^{-1}G_0-(\pi N(0))^2 
\end{align}
\noindent Putting everything together, we find

\begin{align}
    &\phantom{=}\dfrac{\sin^2(k_F R_i)}{(k_F R_i)^2}e^{-R_i/\ell}\dfrac{(\pi N(0))^2}{V_0^{-2}-2V_0^{-1}G_0-(\pi N(0))^2}\bigg\{-V_0^{-1}\dfrac{\bar{\omega}^2+1}{\bar{\omega}^2-1}+G_0\bigg\}\notag\\
    % &=\dfrac{\sin^2(k_F R_i)}{(k_F R_i)^2}e^{-R_i/\ell}\dfrac{1}{(\pi N(0)V_0)^{-2}-2V_0^{-1}(\pi N(0))^{-2}G_0-1}\bigg\{-V_0^{-1}\dfrac{\bar{\omega}^2+1}{\bar{\omega}^2-1}+G_0\bigg\}\notag\\
    % &=\dfrac{\sin^2(k_F R_i)}{(k_F R_i)^2}e^{-R_i/\ell}\dfrac{1}{(\pi N(0)V_0)^{-2}-2V_0(\pi N(0)V_0)^{-2}G_0-1}\bigg\{-V_0^{-1}\dfrac{\bar{\omega}^2+1}{\bar{\omega}^2-1}+G_0\bigg\}\notag\\
    % &=\dfrac{\sin^2(k_F R_i)}{(k_F R_i)^2}e^{-R_i/\ell}\dfrac{1}{(\pi N(0)V_0)^{-2}\{1-2V_0G_0\}-1}\bigg\{-V_0^{-1}\dfrac{\bar{\omega}^2+1}{\bar{\omega}^2-1}+G_0\bigg\}\notag\\
    % &=\dfrac{\sin^2(k_F R_i)}{(k_F R_i)^2}e^{-R_i/\ell}\dfrac{(\pi N(0)V_0)^{2}}{1-2V_0G_0-(\pi N(0)V_0)^{2}}\bigg\{-V_0^{-1}\dfrac{\bar{\omega}^2+1}{\bar{\omega}^2-1}+G_0\bigg\}\notag\\
    % &=N(0)\dfrac{\sin^2(k_F R_i)}{(k_F R_i)^2}e^{-R_i/\ell}\dfrac{(\pi \lambda_V)^2}{1-2V_0G_0-(\pi \lambda_V)^2}\bigg\{-\lambda_V^{-1}\dfrac{\bar{\omega}^2+1}{\bar{\omega}^2-1}+G_0/N(0)\bigg\}\notag\\
    &=N(0)(\pi \lambda_V)^2\dfrac{\sin^2(k_F R_i)}{(k_F R_i)^2}e^{-R_i/\ell}\dfrac{1}{1-2V_0G_0-(\pi \lambda_V)^2}\bigg\{-\lambda_V^{-1}\dfrac{\bar{\omega}^2+1}{\bar{\omega}^2-1}+G_0/N(0)\bigg\}
\end{align}
\noindent The local density of states then becomes

\begin{align}
N(\omega;\,{\bf R}_i)=N_{\textrm{hom}}(\omega)\bigg\{1-\dfrac{\sin^2(k_F R_i)}{(k_F R_i)^2}e^{-R_i/\ell}\dfrac{(\pi \lambda_V)^{2} }{\bigg\{(\pi \lambda_V)^2-1\bigg\}^2 + \dfrac{(2\pi \lambda_V \omega)^{2}}{\omega^2-\Delta^2} } \bigg[(\pi \lambda_V)^{2}+\dfrac{\omega^2+3\Delta^2}{\omega^2-\Delta^2}\bigg]\bigg\}
\end{align}

\noindent We can greatly simplify this, by looking at the multiplicative value and letting $g_0\equiv \pi \lambda_V$:
\begin{align}
    \dfrac{g_0^2\bigg(g_0^2+\dfrac{\omega^2+3\Delta^2}{\omega^2-\Delta^2}\bigg)}{(g_0^2-1)^2+(2g_0)^2\dfrac{\omega^2}{\omega^2-\Delta^2}}=\dfrac{g_0^2}{g_0^2+1}\cdot\frac{{\omega^2}/{\Delta^2}-1+ \dfrac{4}{g_0^2+1}}{{\omega^2}/{\Delta^2}-1+\dfrac{4}{g_0^2+1}\cdot\dfrac{g_0^2}{g_0^2+1}}
\end{align}

\noindent We can therefore write the LDOS as 

\begin{align}
N(\omega;\,{\bf R}_i)=N_{\textrm{hom}}(\omega)\left\{1-\dfrac{\sin^2(k_F R_i)}{(k_F R_i)^2}e^{-R_i/\ell}\dfrac{g_0^2}{g_0^2+1}\cdot \dfrac{\omega^2/\Delta^2-1+\dfrac{4}{g_0^2+1}}{\omega^2/\Delta^2-1+\dfrac{4}{g_0^2+1}\cdot \dfrac{g_0^2}{g_0^2+1}}\right\}
\end{align}
\noindent which reproduces what we had in the main text.

\bibliography{main}{}